\documentclass[preprint]{aastex}
\usepackage{subfigure}
\usepackage{enumerate}
\usepackage{epsfig}
\usepackage{lscape}
\usepackage{color,soul}
\usepackage[normalem]{ulem}

\pdfoutput=1

\begin{document}

\title{A Comparison of Spectroscopic versus Imaging Techniques for Detecting Close Companions to \textit{Kepler} Objects of Interest}


\author{Johanna K. Teske\altaffilmark{1, 10, +}, Mark
  E. Everett\altaffilmark{2, 9, 10}, Lea
  Hirsch\altaffilmark{3, *}, Elise Furlan\altaffilmark{4}, Elliott
  P. Horch\altaffilmark{5, 6, 9, 10},
  Steve B. Howell\altaffilmark{7, 9, 10}, David
  R. Ciardi\altaffilmark{4, 9, 10}, Erica
  Gonzales\altaffilmark{8}, Justin R. Crepp\altaffilmark{8}}

\altaffiltext{1}{Carnegie DTM, 5241 Broad Branch Road, NW, Washington,
  DC 20015, USA, email: jteske@carnegiescience.edu}
\altaffiltext{2}{National Optical Astronomy Observatory, 950 N. Cherry Ave, Tucson, AZ 85719, USA}
\altaffiltext{3}{Astronomy Department, University of California at
  Berkeley, Berkeley, CA 94720, USA}
\altaffiltext{4}{NASA Exoplanet Science Institute, California
  Institute of Technology, 770 South Wilson Ave., Pasadena, CA 91125, USA}
\altaffiltext{5}{Department of Physics, Southern Connecticut State University, 501 Crescent Street, New Haven, CT 06515, USA}
\altaffiltext{6}{Adjunct Astronomer, Lowell Observatory, 1400 W Mars
  Hill Rd, Flagstaff, AZ 86001, USA}
\altaffiltext{7}{NASA Ames Research Center, Moffett Field, CA 94035, USA}
\altaffiltext{8}{Department of Physics, University of Notre Dame, 225 Nieuwland Science Hall, Notre Dame, IN 46556, USA}
\altaffiltext{9}{Visiting Astronomer, Kitt Peak National Observatory, National Optical
Astronomy Observatory, which is operated by the Association of Universities
for Research in Astronomy (AURA) under cooperative agreement with the
National Science Foundation, USA.}
\altaffiltext{10}{Visiting Astronomer, Gemini Observatory, National Optical Astronomy
Observatory, which is operated by the Association of Universities for Research
in Astronomy, Inc., under a cooperative agreement with the NSF on behalf of
the Gemini partnership: the National Science Foundation (United States), the
Science and Technology Facilities Council (United Kingdom), the National
Research Council (Canada), CONICYT (Chile), the Australian Research
Council (Australia), inist\'{e}rio da Ci\^{e}ncia, Tecnologia e Inova\c{c}\~{a}o 
(Brazil) and Ministerio de Ciencia, Tecnolog\'{i}a e Innovaci\'{o}n Productiva (Argentina).}

\altaffiltext{+}{Carnegie Origins Fellow, jointly appointed by Carnegie DTM \& Carnegie Observatories}
\altaffiltext{*}{IPAC/Caltech Visiting Graduate Fellow}

\clearpage

\begin{abstract}

\textit{Kepler} planet
candidates require both spectroscopic and imaging follow-up observations
to rule out false positives and detect blended stars.
Traditionally, spectroscopy and high-resolution imaging 
have probed different host star companion parameter spaces, the former detecting
tight binaries and the latter detecting wider bound companions as well as chance background
stars. In this paper, we examine a sample of eleven \textit{Kepler} host
stars with companions detected by two techniques -- 
near-infrared adaptive optics and/or optical speckle interferometry imaging, and a
new spectroscopic deblending method. We compare the companion
 effective temperatures (T$_{eff}$) and flux ratios (F$_B$/F$_A$, where $A$ is
the primary and $B$ is the companion) derived from each technique, and
find no cases where both companion parameters agree within 1$\sigma$ errors. In 3/11 cases the companion T$_{eff}$ values agree within 1$\sigma$ errors,
and in 2/11 cases the companion F$_B$/F$_A$ values agree within 1$\sigma$
errors. Examining each \textit{Kepler} system individually considering
multiple avenues (isochrone mapping, contrast curves, probability of
being bound), we suggest two cases for which the techniques most likely agree in their companion detections (detect the same
companion star). Overall, our results support the advantage that spectroscopic deblending
technique has for finding very close-in companions
($\theta \lesssim$0.02-0.05$\arcsec$) that are not easily detectable with
imaging. However, we also specifically show how high-contrast AO
and speckle imaging observations detect companions at larger
separations ($\theta \geq$0.02-0.05$\arcsec$) that are missed by the spectroscopic technique, provide
additional information for characterizing the companion and its
potential contamination (e.g., position angle, separation, magnitude
differences), and cover a wider range of primary star effective temperatures. 
The investigation presented here illustrates the utility of combining the two techniques
to reveal higher-order multiples in known planet-hosting systems.

\end{abstract}


\section{Introduction}


Bound companions to exoplanet host stars may influence the planet formation
and evolution process in multiple ways, from the very first stages of
planet ``birth'' to after planets have fully formed and are
interacting with each other/other stars: truncation and dynamical
heating of the protoplanetary disk (e.g. Artymowicz \& Lubow 1994;
Mayer et al. 2005; Pichardo et al. 2005; Kraus et al. 2012), ejection
of planets (e.g., Kaib et al. 2013; Zuckerman 2014), and migration of
planets (e.g., Wu \& Murray 2003; Fabrycky \& Tremaine 2007; Naoz et
al. 2012). 
Despite simulations predicting that both tight and
wide bound companions to host stars can hinder planet formation (e.g., Bouwman et al. 2006;
Fabrycky \& Tremaine 2007; Jang-Condell 2007; Thebault 2011;
Malmberg et al. 2011; Kaib et al. 2013; Petrovich 2015), numerous
exoplanets have been detected in binary/multiple star systems (e.g.,
Eggenberger et al. 2007; Raghavan et al. 2010; Orosz et al. 2012a,
2012b), including circumbinary planets (e.g., Doyle et al. 2011; Welsh
et al. 2012; Orosz et al. 2012a, b; Schwamb et al. 2013, Kostov et
al. 2014). Thus binarity plays a role in planet formation, but does not strictly preclude it.

 
The explosion of exoplanet targets found by \textit{Kepler} has
allowed for more thorough investigations of host star binarity using
the \textit{Kepler} sample, which is not influenced by the selection
bias of radial velocity planet detection surveys (e.g., small separation
binaries are avoided in radial velocity planet searches). These studies
indicate that planet formation is suppressed in multiple-star systems
with separations $\lesssim$1500 AU (Wang et al. 2014), and
hint that stellar multiplicity affects different types of planet
formation in different ways (Ngo et al. 2015; Wang et
al. 2015). Notably, Horch et al. (2014) (H14) combined the measured
detection limits from high-resolution
speckle imaging observations of over 600 \textit{Kepler} Objects of
Interest (KOIs) -- stars that show potential planetary object signatures in their
light curves -- with statistical properties of known binary systems
and a model of the Galactic stellar distribution (TRILEGAL; Girardi et
al. 2005) to estimate how many exoplanet host stars in the
\textit{Kepler} field of view are in spatially-resolvable binary systems. Their simulation
predicts that most of the sub-arsecond companions detected around
\textit{Kepler} stars with imaging are physically bound to the primary
star, meaning that in general (over the separation range that such observations
are sensitive, $\sim$0.1-1$\arcsec$) exoplanet host stars have a binary
fraction similar to that of field stars, $\sim$40-50\%. 

In addition to characterizing the binarity of hosts to exoplanets to
learn more about how planets form,
particularly small planets like those found by \textit{Kepler}, detecting companions to
\textit{Kepler} host stars is important for measuring accurate radii
of the planets themselves. It is only with accurate and precise (to $\sim$20\%; e.g.,
Rogers 2014) radii (and thus density) measurements that we can
distinguish between ``rocky''/terrestrial and not-rocky/not-Earth-like
planets. The large pixel size ($\sim$4$\arcsec$$\times$4$\arcsec$), aperture, and
centroiding algorithm of the \textit{Kepler} pipeline still allow
for false positives and blended stars to introduce dilution to transit
measurements, resulting in underestimates of planetary radii and
overstimates of planetary density. This issue spurred a
dedicated and expansive community follow-up program to the space-based
\textit{Kepler} observations, including spectroscopy and
high-resolution imaging to detect false positives and close
companions, to confirm the exoplanets and refine their host
star parameters (e.g., Howell et al. 2011; Horch et al. 2012; Horch et
al. 2014; Everett et al. 2015; Ciardi et al. 2015). Most recently,
Ciardi et al. (2015) fit isochrones, based on the stellar parameters
from the NASA Exoplanet Archive, to each potential host star in the
cumulative \textit{Kepler} candidate list and calculated for each the possible
factor by which the orbiting planet radii are underestimated, assuming
five different multiple-star scenarios (e.g., planet orbits primary
star, planet orbits secondary star, planet orbits tertiary star,
etc.). Their resulting radius correction factor -- the degree to which
planetary radii are underestimated based on the presence of undeteted
stars -- varies for each system
and for each multiplicity scenario, but the overall mean correction
factor for stars observed by \textit{Kepler} with no follow-up
observations is 1.49$\pm$0.12.
The mean correction factor for
stars with typical follow-up observations (2-3 radial velocity
measurements over 6-9 months, spectroscopy of the primary star, and
high resolution imaging in at least one filter) is reduced to
1.2$\pm$0.06, illustrating how crucial such observations are to
understanding the basic characteristics of detected planets.

Traditionally, spectroscopy and imaging follow-up of KOIs have been
used to probe different parameter spaces of companions,  spectroscopy
being important for detecting tight binaries and imaging being more
relevant for wider bound companions as well as chance background
stars. Recently, Kolbl et al. (2015; K15) introduced a new technique for
detecting close companions to KOIs using \textit{Keck/HIRES} spectroscopy
originally purposed for measuring radial velocities of planets (and
thus their masses) and/or close stellar companions. In this paper, we
aim to refine the answer to the question,  
\textit{Do the spectroscopic and imaging techniques detect the same or
  different stars?} Specifically, we compare
the properties of companions detected by K15 using spectroscopy and
companions detected by various high-contrast imaging \textit{Kepler}
follow-up campaigns, to see whether the two methods overlap in their
detection rates and characterization of detected companions. 

\section{Data Examined In This Work}

\subsection{HIRES Spectroscopic Detections of Companions}

K15 present a method for detecting close companions to KOIs,
many of which host planet candidates that still require
validation. K15 search through the California \textit{Kepler}
Survey's catalog of 1160 single-epoch, high resolution optical \textit{Keck I/HIRES}
spectra of KOIs for evidence of more than one set of stellar
absorption lines. They systematically test whether each individual spectrum is
best represented as the sum of two or more
input spectra drawn from an extensive library of model spectra
spanning the H-R diagram. From their analysis, K15 detect companions
to 63 KOIs, and provide effective
temperatures (T$_{eff}$) and flux ratios (F$_B$/F$_A$, where $A$ is
the primary and $B$ is the companion) measured across the $V+R$ bands for each companion. 

As with any detection method of close-in companions to stars, K15's
spectroscopic ``contamination'' detection method has caveats
that are described in more detail in that work. Briefly, K15 are only
sensitive to companions that fall within the slit, which corresponds
to distances 
0.43-1.5$\arcsec$ from the primary star. 
They assume in their model-fitting
process that the primary star is on the Main Sequence, and their model
templates fall between 3200 and 6500 K. Their library of companion star
templates contains members between 3300 and 6100 K, but it does not contain a
representative median spectrum between 3800 and 3900 K. The K15 code
cannot detect companion stars with $\Delta$RVs -- the relative radial
velocity between primary star and the potential companion(s) -- less
than 10 km/s, and is limited to companions with orbital
periods $\gtrsim$2.5 days, corresponding to the maximum detectable
Doppler shift of $\pm$200 km/s. (There is an
exception for M dwarfs orbiting G-type primary stars, explained below.) Furthermore, if the primary and
companion spectral types are similar and their relative RV is low
($\lesssim$ 20 km/s), the flux of the companion star can be
underestimated if some of its flux is subtracted away with that of the
primary star in the K15 analysis. This can in turn decrease the
calculated flux ratio for the two stars. 

In general, the K15 method is
able to detect companion stars with as small a spectral contribution as
0.5\%-1\% of primary star's flux. Their method is most accurate for
companions with a $\Delta$RV $>$ 10 km/s, a $<$20\% flux contribution,
and when both primary and companion stars have 3000 K$<$T$_{eff}
<$6000 K. 
Their injection-recovery tests that paired
actual spectra of their sample with the designated companion star shifted by $\Delta$RV +50
kms$^{-1}$ indicate a
range of recovery rates depending on primary vs. companion T$_{eff}$
and \% of total flux contributed by the companion (see their Table 2). 
For the
synthetic binary cases where the K15 algorithm recovered the companion
star, its predicted temperature and flux ratio in the visible also
vary in their accuracy (see their Table 5), e.g., the largest
$\sigma{T_{eff}}$ (950 K) occurs in a 3500 K primary+6000 K companion system
in which the companion contributes 1\% of the total flux, and the
largest $\sigma_f/f$ (0.55) occurs in both a 5500 K primary+5500 K
companion system and a 3500 K primary+5000K companion system, in which the companion contributes 1\% of the total
flux. 

K15 test more thoroughly cases of G-type primaries
(5500 K)+ M-dwarf companions (3500 K) with $\Delta$RV$=$5 kms$^{-1}$
and companion flux contributions of 1\%, 3\%, or 5\% of the total
flux. The recovery rate for companion stars with 3\% or 5\% of the total
flux is 90\%, whereas the recovery rate for companions with 1\%
of the total flux drops to 40\%. Similarly,  the deduced T$_{eff}$
differs more from the actual T$_{eff}$ as the flux contribution of the
companion star decreases, although the deduced versus actual percentage
flux decreases with decreasing \% flux of the companion (see their Table 6). When the M dwarf contributes 0.5\% of
the total flux, at a $\Delta$RV of 50 kms$^{-1}$, the detection rate
is also 40\%, but no fainter companions (at 0.05\% or 0.1\% of the
total flux) are recovered. 

In \S4.2.1 and \S4.2.2, we examine how different limitations of the observations an analysis between the spectroscopic detection methods and the imaging detections methods (which will be discussed in the next section) influence the derived companion parameters.

\setcounter{footnote}{11}
\subsection{New Imaging Observations}
We aim to compare the properties of companions detected via the
K15 spectroscopic method to those detected by speckle interferometry
and adaptive optics (AO) imaging. In Table \ref{tab1} we list all of the KOIs with companions detected
by K15 that also have companions detected in imaging data available through the
Community Follow-Up Observing Program (CFOP)\footnote{https://cfop.ipac.caltech.edu/}, an online public
repository for observations and measured properties of KOIs, or from our own
observations. Below we detail the new
observations and data reduction that have not been
previously published. 

\subsubsection{NIR AO Observations}
Near-infrared adaptive optics imaging was acquired at Palomar Observatory 
for KOIs 1613, 3161, and 3471 and at \textit{Keck} Observatory for KOIs 
5, 652, 1361, 1613, and 2311. The Palomar observations utilized PHARO 
(Hayward et al. 2001) on the Hale 5m telescope and the \textit{Keck} observations 
utilized NIRC2 (Wizinowich et al. 2004) on the 10-m \textit{Keck} II telescope; the 
observations at each telescope were made in different near-IR filters
(see Table \ref{tab1}). 
Observations at both telescopes utilized the adaptive optics system,
with each target as a natural guide star. At Palomar a 5-point quincunx dither 
pattern was used, and at \textit{Keck} a three-point dither pattern was used,
to avoid the lower left quadrant of the NIRC2 array. Three images were collected at 
each dither pattern position, each shifted 0.5$\arcsec$ from the previous 
dither pattern. On Palomar, PHARO has a field of view of 25$\arcsec$$\times$25$\arcsec$ with 
a pixel scale of 25 mas; the dither size was 5$\arcsec$ yielding a final 
coadded final field of approximately 10$\arcsec$. On \textit{Keck}-II, NIRC2 has a 
field of view of 10$\arcsec$$\times$10$\arcsec$ with a pixel scale of 10 mas; with a dither 
pattern size of 2$\arcsec$, the final coadded field of view was 
approximately 4$\arcsec$x4$\arcsec$.

Sky frames were constructed for each target from the target frames 
themselves by median filtering and coadding the dithered frames. Each 
dither pattern frame was then sky subtracted and flatfielded. Individual 
exposures per frame varied depending on the brightness of the target but 
typically were 10 -- 30 seconds per frame, yielding photometry 
on the primary target from the final coadded dither pattern of $S/N 
\gtrsim 500$. Data reduction was performed with a custom set of IDL 
routines.

Aperture photometry was used to obtain the relative magnitudes of stars 
for those fields with multiple sources. Point source detection limits 
were estimated in a series of concentric annuli drawn around the star. 
The separation and widths of the annuli were set to the FWHM of the 
primary target point spread function. The standard deviation of the 
background counts is calculated for each annulus, and the $5-\sigma$ 
limits are determined within annular rings (see also Adams et al. 2012). 
The PSF widths for the Palomar and \textit{Keck} images were typically found to be 
4 pixels for the two instruments corresponding to 0.1$\arcsec$ and 0.04$\arcsec$
FWHM, respectively. Typical contrast levels are 2 –- 3 mag at a
separation of 1 FWHM and 7 -– 8 mag at $>$5 FWHM with potentially deeper 
limits past 10 FHWM. 
We did not detect a companion for KOI 3471, but
its FWHM in the Palomar/PHARO image is very large, about 0.35", and thus we
cannot exclude the presence of a close companion.

\subsubsection{Optical Speckle Observations}


Both KOI 5 and KOI 1613 were observed using the Differential Speckle
Survey Instrument (DSSI; Horch et al. 2009) at the \textit{WIYN} 3.5-m 
Telescope at Kitt Peak in a set of two optical filters (see Table
\ref{tab1}). DSSI is composed of two 512x512 16 $\mu$m EMCCDs attached
at perpendicular ports; the light from the telescope is split by a
dichroic to go through two filters to two cameras simultaneously. At
\textit{WIYN}, the DSSI plate scale was measured to be 0.0217 and
0.0228 arcsec pixel$^{-1}$ for the two cameras,  resulting in a $\sim$11.1x11.7
arsec$^{2}$ field of view, although often only a subregion of the EMCCDs are read
out to expedite observations (e.g., 128x128, $\sim$2.8x2.9 arcsec$^{2}$
FOV) (Howell et al. 2011). DSSI is diffraction-limited, which at \textit{WIYN} gives a resolution
of $\sim$0.05$\arcsec$. In the case of KOI 1613, observations were 
initially made on 13 June 2011 and then further data were obtained 
on 21 and 23 September 2013. For KOI 5, three observations have also
occurred: on 17 September 2010, 18 September 2010, and 21 September 2010. The number of 
speckle data frames obtained in each filter for these observations was 
between 3 and 5 thousand for KOI 5 and 1 to 4 thousand for KOI 1613.
The frame exposure time was 40 ms in all cases. The data are stored 
in 1000-frame FITS files, and the results from multiple files on a given 
star were coadded to obtain the final result.

A full description of the method for the data reduction and analysis
for \textit{WIYN} DSSI data has been given in e.g. Horch et al. (2011) and 
Howell et al. (2011). However, a brief description is warranted here.
From the raw data, we form the autocorrelation and triple correlation of
each data frame, sum these over the entire frame sequence, and then 
Fourier transform these to obtain the total spatial frequency power 
spectrum and total spatial frequency bispectrum of the observation.
To calculate a reconstructed image of the target, we deconvolve the
power spectrum with that of a bright unresolved star observed in the same
way, close in time and in sky position to that of the binary star.
(In the Fourier domain, the deconvolution is performed by dividing the
power spectrum of the binary with that of the point source.) Taking
the square root of this function, we obtain the modulus of the object's
Fourier transform. On the other hand, the bispectrum contains information
that allows for the calculation of the phase of the object's Fourier
transform, which we estimated using the method of Meng et al. (1990). The modulus and phase are combined, the result is low-pass
filtered to suppress noise above the diffraction limit, and then it is
inverse-Fourier transformed to arrive at the reconstructed image.
The same process is used for the data stacks in both filters resulting in
reconstructed images for each color.

The reconstructed images are then visually inspected for companions. If a 
companion is found in both images (which is the case for the objects here),
then the approximate position relative to the primary star is noted 
from the reconstructed image, and used as the starting position for
a downhill simplex fitting routine to obtain the final differential
astrometry and photometry of the system. However, the fitting is done to 
the deconvolved power spectrum, where the signature of a companion is 
a fringe pattern (i.e. a cosine squared function). The spacing, orientation, 
and fringe depth are uniquely determined by the separation, position 
angle, and magnitude difference of the binary star.


DSSI was used at the \textit{Gemini} North Telescope in July
of 2014. At \textit{Gemini}, the plate scale is 0.011 arcsec pixel$^{-1}$, and
often the camera is windowed to a smaller pixel region (e.g., 256x256,
or 2.8x2.8$\arcsec$). The diffraction-limited resolution at \textit{Gemini} is 0.016$\arcsec$
at 500nm and 0.025$\arcsec$ at 800nm (Horch et al. 2012). We obtained observations of KOI 2059 and KOI 3471 among
a large number of Kepler Objects of Interest. KOI 2059 was 
observed on two dates on that run, namely 19 July and 24 July while
KOI 3471 was only observed once, on 24 July. The data collection was similar
to \textit{WIYN} observations in terms of number of frames and data collection
in 1000-frame subsets, but the frame exposure time used at \textit{Gemini}
was 60 ms, which is longer that that used at \textit{WIYN} owing to the 
better average seeing conditions at \textit{Gemini} versus \textit{WIYN}. This means that
the correlation time of the atmosphere is longer at \textit{Gemini}, and
therefore the speckle lifetimes on the image plane are also longer. 

We also store the data from \textit{Gemini} in larger arrays than \textit{WIYN}, 256x256 
pixel frames for \textit{Gemini} versus 128x128 pixels for \textit{WIYN}. This is needed
since the magnification of the images is higher than what we use at 
\textit{WIYN} in order to sample the (diffraction-limited) speckles properly
at the larger aperture. However, once the data are collected, the
reduction steps are identical to what is described above for the 
\textit{WIYN} observations. More details on \textit{Gemini} speckle data reductions with
DSSI data can be found in Horch et al. (2012).

\section{Analysis}


The overall sample of KOIs with spectroscopy versus the sample with imaging observations
is different -- the former is generally limited to brighter stars
(and is rather complete at Kepler magnitudes $<$14.2), since the
observations necessarily disperse the light -- and the techniques
provide different information, e.g., astrometry can
only be derived from imaging observations. For many KOIs, only one type of observation is
available, so it is important to understand how this limitation
affects the detection rate and characterization of close companions. 
Here we examine a unique sample of eleven KOIs that have companions
detected from both the spectroscopic deblending method of K15 and
NIR/AO and/or speckle imaging. Our goal is to determine whether the companions detected by K15 are the same companions detected by the imaging efforts, or whether each technique uncovers a completely separate sample of companion stars.

K15 report T$_{eff}$ values and flux ratios in $V+R$ based on their
spectroscopic analysis for each of the KOI companions they detect. The
directly-measured quantities from imaging data are separation,
position angle, and $\Delta$m ($\Delta$magnitude). Combining the measured 
$\Delta$m values (and when available, multiple colors) of the
imaging-detected companions with the known properties of the 
primary KOIs, we attempt to derive companion T$_{eff}$ values and primary-to-companion
flux ratios for each of the companions to KOIs listed in Table 1. 

Using differential photometry, it is possible to calculate companion
effective temperatures via isochrone fitting techniques (specifically,
by shifting primary star properties down an isochrone to derive
companion parameters). However, this analysis relies on the assumption
that the companion star is physically bound, and should lie along the
same isochrone as the primary star. This assumption can only be
assessed in the case of multi-filter photometry, with which we can
construct a color-magnitude diagram and empirically test whether the
two stars are consistent with the same set of isochrones. In the
following sections, we describe the isochrone-fitting process by which
we attempt to determine whether the observed imaging companions are
gravitationally bound, and thus whether their derived effective
temperatures are accurate. Of the 11 stars with companions detected by
both K15 and high-resolution imaging, 7 have multi-filter photometry.

\subsection{Properties of Detected Companions Derived from Imaging Data}

For each of the 11 KOIs with detected companions in K15 and at
least one high-resolution imaging detection of a companion, we use the
isochrone fitting procedure of Everett et al. (2015) to map out the
photometric probability distribution of the primary star based on the
Dartmouth isochrones. We use as inputs the primary star's inferred
$T_{eff} $, log($g$), and [Fe/H] from Huber et al. (2014), as listed
in Table \ref{tab0}. For each
mass point on a set of isochrones ranging in age from 1-13 Gyr (at 0.5
Gyr intervals) and metallicity from -2.5 to $+$0.5 (in 0.02 dex
intervals), we assign a probability value between 0 and 1 based on its
proximity to the input stellar parameters. Since each mass point on an
isochrone is associated with a set of absolute magnitudes in various
filters, we can convert the primary star's isochrone probability
distribution in $T_{eff}$-log($g$)-[Fe/H] parameter space to an
absolute magnitude likelihood distribution in each of the filters in which we have photometric data. This allows us to plot the primary star's probability distribution in color-magnitude-metallicity space.

\begin{deluxetable}{lcccc}
\tabletypesize{\scriptsize}
\tablecolumns{5}
\tablewidth{0pc}
\tablecaption{Primary KOI Stellar Parameters from Huber et al. (2014)}
\tablehead{ 
\colhead{KOI} &\colhead{KIC ID} & \colhead{T$_{eff}$ (K)} &\colhead{log $g$ (dex)} & \colhead{[Fe/H] (dex)} }
\startdata
5 	&   8554498  & 5753$\pm$115 & 4.003$\pm$0.03 & 0.05$\pm$0.15 \\
652	 &   5796675  & 4694$\pm$137 & 4.791$\pm$0.40 & -1.45$\pm$0.03 \\
1152 & 10287248 & 3806$\pm$80 & 4.773$\pm$0.15 & -0.13$\pm$0.15 \\
1361 &6960913 & 4017$\pm$80 & 4.656$\pm$0.40 & 0.03$\pm$0.15\\
1452 &7449844 & 7162$\pm$240&4.100$\pm$0.40 & -0.18$\pm$0.30 \\
1613 &6268648 & 6044$\pm$120& 4.192$\pm$0.03 &-0.24$\pm$0.15\\
2059 &12301181 & 4997$\pm$99 & 4.597$\pm$0.15 & -0.01$\pm$0.15\\
2311 &4247991& 5765$\pm$115 & 4.720$\pm$0.15 & 0.17$\pm$0.15\\
2813 &11197853& 5133$\pm$151&4.237$\pm$0.40 & -0.99$\pm$0.30\\
3161 &2696703 & 6795$\pm$237 & 4.182$\pm$0.40 & 0.18$\pm$0.30\\
3471 &11875511 & 4821$\pm$135 & 3.787$\pm$0.40 & 0.01$\pm$0.30\\
\enddata
\label{tab0}
\end{deluxetable}

The shapes of the resultant primary star probability distributions are determined both by the size of the error bars on the input parameters ($T_{eff}$, log($g$), and [Fe/H]) and on the available parameter space covered by the set of Dartmouth isochrones we use. In the case of large uncertainty in the input parameters, the shape of the distribution may be truncated by the allowable parameter space from the available isochrones.


By combining the primary star isochrone fits with 
photometric data of each KOI's companion taken in different bands, we can derive a probability
distribution for the companion, in magnitude and physical parameter
space, by assuming the secondary star falls on the same isochrone as
its primary. In other words, we assume it is a bound companion. For
each filter in which we measure a $\Delta$m, we shift each
isochrone mass point (and associated probability level) down its
respective isochrone according to the differential photometry. This
produces an ``isochrone-shifted'' stellar parameter distribution for the companion star,
indicating its T$_{eff}$ and log($g$) as well as its absolute
magnitudes in various filters, assuming it shares a metallicity and
age with its primary KOI. The
resulting {\it bound} companion parameters -- T$_{eff}$ and flux
ratio, converted to the \textit{Kepler K$_p$} bandpass -- are listed
in Table \ref{tab2}. The K15 flux ratios are based on data
covering $\sim$500-800 nm, whereas the K$_p$ bandpass is $\sim$400-900
nm. To check that the K$_p$ flux ratios we derived for each
KOI+companion pair did not differ significantly
from the same flux ratio calculated with our derived V+R
magnitudes, we used the V and R magnitudes to calculate new flux ratios, using the zero point offsets of Bessell et al. (1998). In every case, the V+R band flux ratio was the same as the Kepler band flux ratio within 1$\sigma$ errors, in most cases within half the error or less.

In cases where there are multi-band photometric observations of the
companion, we can assess the assumption that it is bound, and
therefore bolster confidence in the properties derived from
the isochrone shifting based on differential photometry. By comparing the
``isochrone-shifted'' probability distributions derived from two or
more distinct $\Delta$m values, we can determine whether the photometry in each individual filter is consistent with the same bound companion star. If photometry in two filters produce companion models that are mutually inconsistent, the assumption of a bound companion is likely false.

To assess the consistency between the isochrone-shifted probability
distributions produced by the various $\delta$m values, we calculate the coefficient of overlapping, $OVL$ 
(Schmid \& Schmidt 2006). This coefficient is designed to measure the common area underneath two 
distributions. We marginalize each probability distribution in metallicity, then 
normalize each distribution to sum to unity. We then calculate the overlap coefficient as
the sum of the minimum value between the two distributions at each position in 
color-magnitude space:
$$OVL = \sum_{color} \sum_{mag} \min(f_{1}(color,mag), f_{2}(color,mag))$$
Here, $f_1$ and $f_2$ represent the marginalized distributions produced by $\Delta$m$_1$ and 
$\Delta$m$_2$, the differential magnitudes in two distinct filters. Each distribution is
a function of color and magnitude, and we sum over the entire range in color and magnitude
covered by both distributions. These $OVL$ values are listed in Table
\ref{ovl}. Since the overlapping coefficient depends strongly on
the specific shapes of the probability distributions, we use it as a
comparative assessment of the similarity between probability
distributions produced by $\Delta$m's in different bands, rather than as
a hard cutoff for bound versus unbound companions.

\begin{deluxetable}{lccc}
\tabletypesize{\scriptsize}
\tablecolumns{4}
\tablewidth{0pc}
\tablecaption{Coefficient of Overlap Between Isochrone-Shifted
  Probability Distributions }
\tablehead{ 
\colhead{KOI} & \colhead{avg (F692-K) $OVL$} & \colhead{avg (F692-F880)
  $OVL$} & \colhead{avg (J-K) $OVL$} }
\startdata
5 	                        &        0.677       & \nodata & \nodata \\
652 (B companion)	 &   \nodata & \nodata &  0.273  \\
652 (C companion)    &  \nodata & \nodata &  0.254  \\
1361                          &  \nodata & \nodata &   0.869 \\
1613                          &         0.378       &     0.845         & \nodata \\
2059                          &     0.006         &  \nodata & \nodata \\
2311                          &     0.00          & \nodata &         0.375   \\
3471  (subgiant)        & \nodata  &     0.399         & \nodata \\          
3471 (dwarf)              & \nodata  &  0.716            & \nodata \\
\enddata
\label{ovl}
\end{deluxetable}

In a similar analysis, the relative magnitudes and colors of any bound
companion, measured with respect to the modeled absolute magnitude and
colors of the primary, should fall on the same isochrone and be
coincident with the isochrone shifted properties.  When this fails,
it provides evidence against the bound assumption.  In cases where
a companion star is deemed unlikely to be a bound companion, its
stellar properties derived using these methods should be considered
invalid.

\begin{figure}[h]
\centering
\includegraphics[width=\textwidth]{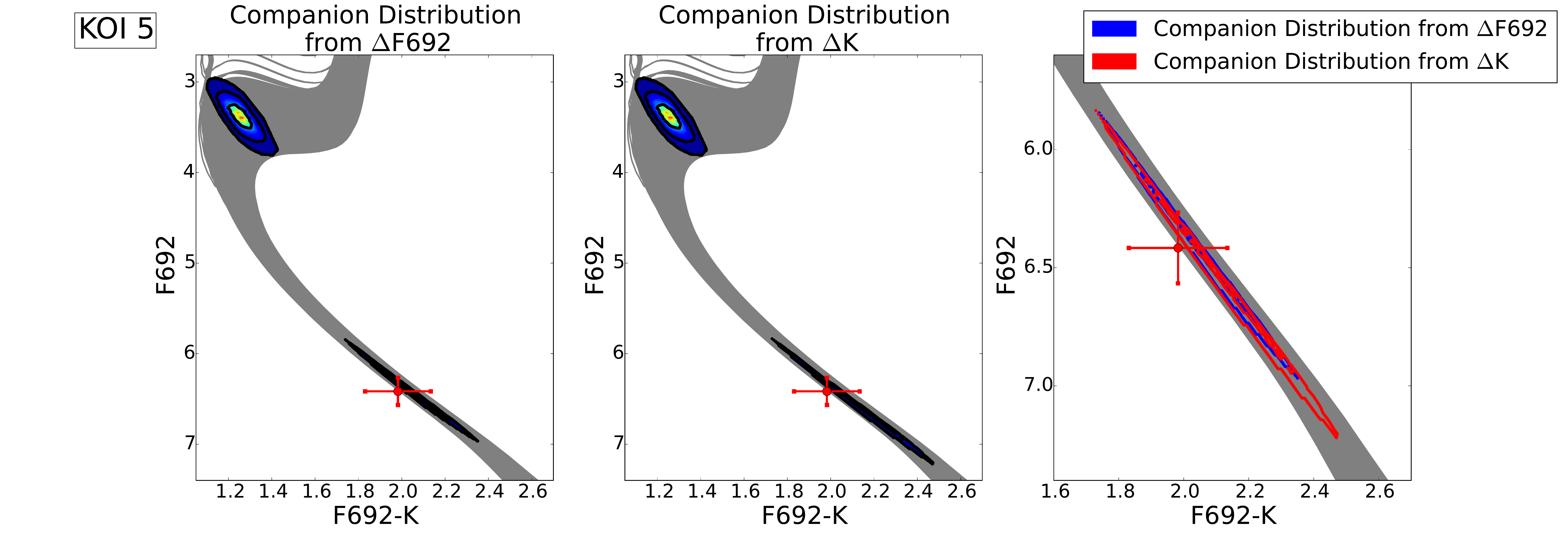}
\quad
\includegraphics[width=\textwidth]{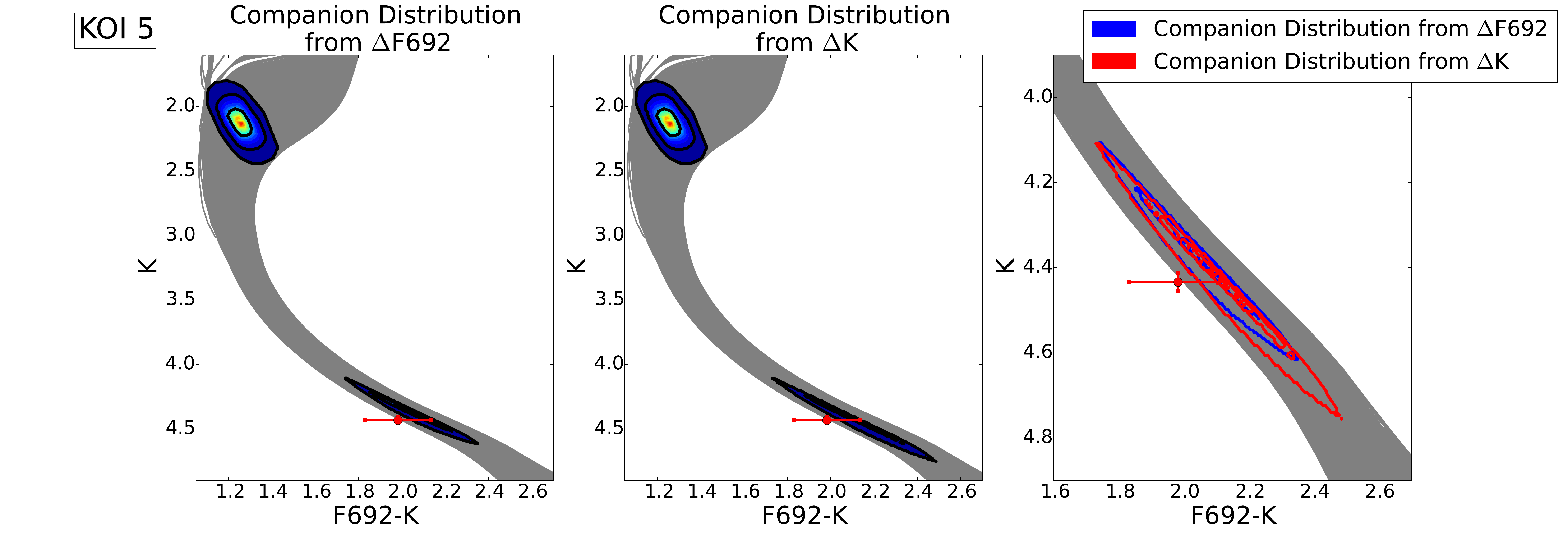}
\caption{Plots showing the results of the imaging data analysis in \S3.1 for
  KOI 5. Left: Primary KOI
  absolute photometry contours, and companion photometry
  contours, calculated from observed $\Delta$F692 magnitude and assuming it lies at the same distance and has
  the same age and metallicity as the KOI, mapped on the same (primary
  KOI) isochrone. The red point represents the absolute magnitude and
  ``true'' color for the companion (assuming it is bound), calculated from
  relative color information. The spread in color of the contours
  represents the spread in the normalized probability distribution,
  ranging from 1 (red) to 0 (dark blue).  Middle: Same as left, but with companion photometry
  contours calculated from $\Delta$K magnitude. Right: A comparison of
  the overlap between the relative photometry contours of the companion. The red point here
is the same as in the left and middle panels.}
\label{iso_koi5}
\end{figure}

\begin{figure}[h]
\centering
\includegraphics[width=0.95\textwidth]{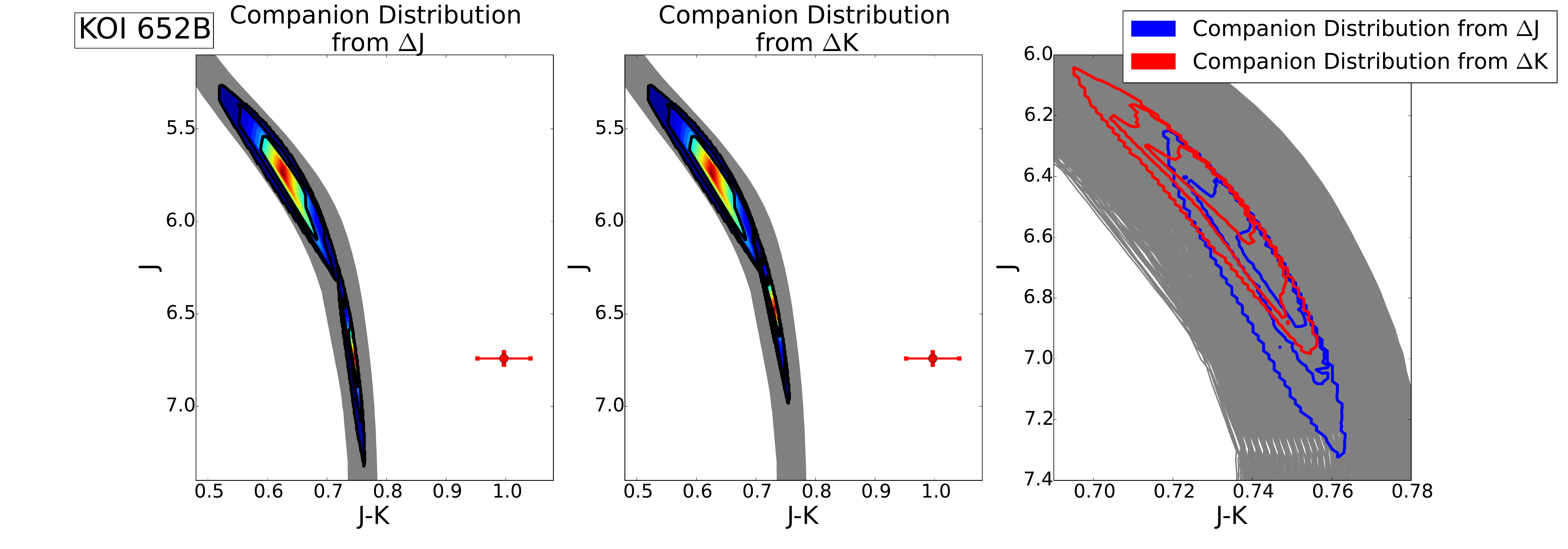}
\quad
\includegraphics[width=0.95\textwidth]{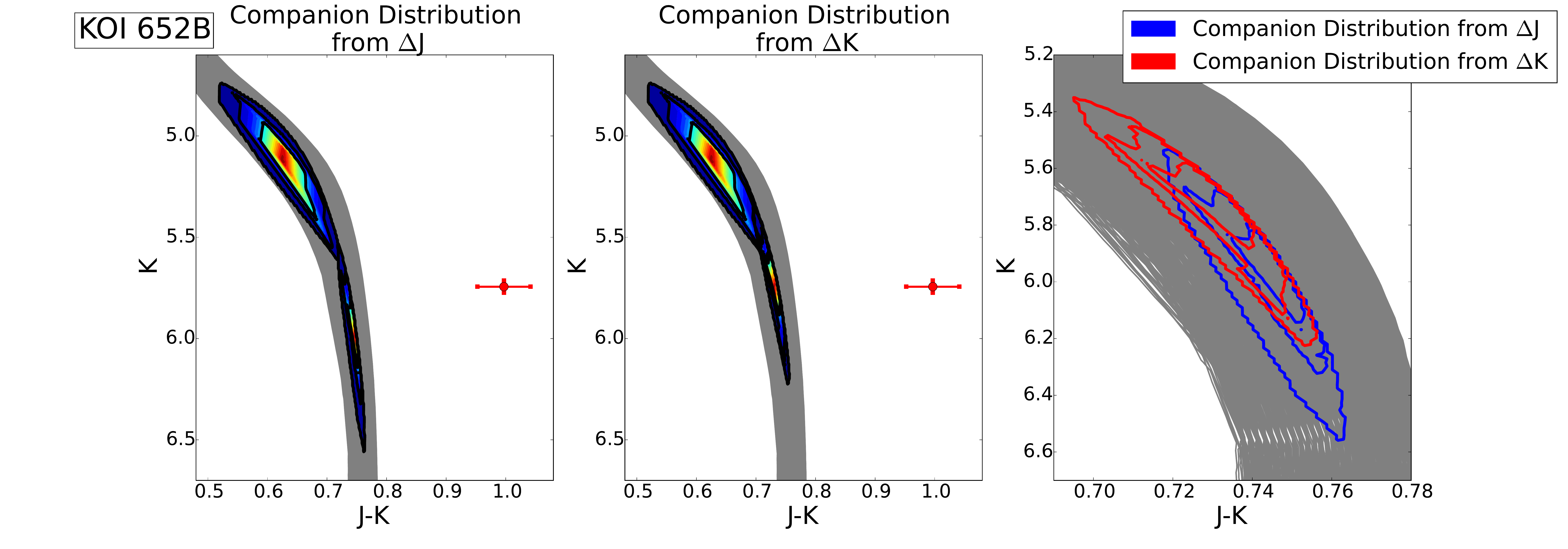}
\quad
\includegraphics[width=0.95\textwidth]{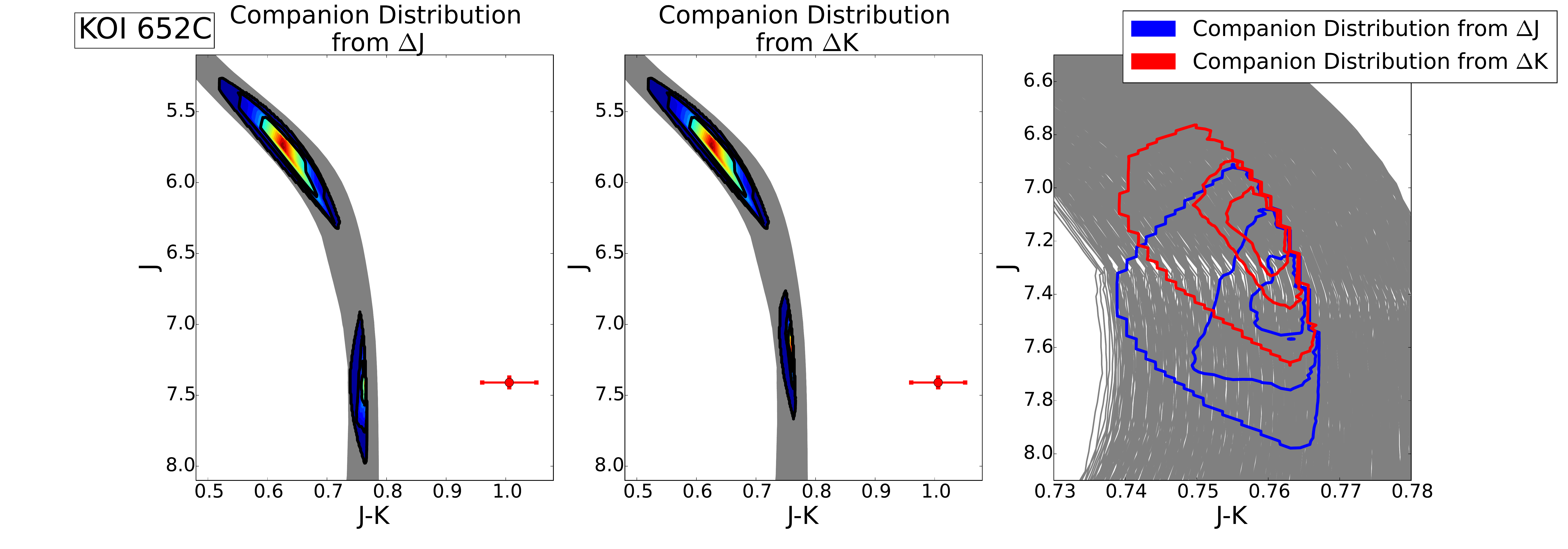}
\quad
\includegraphics[width=0.95\textwidth]{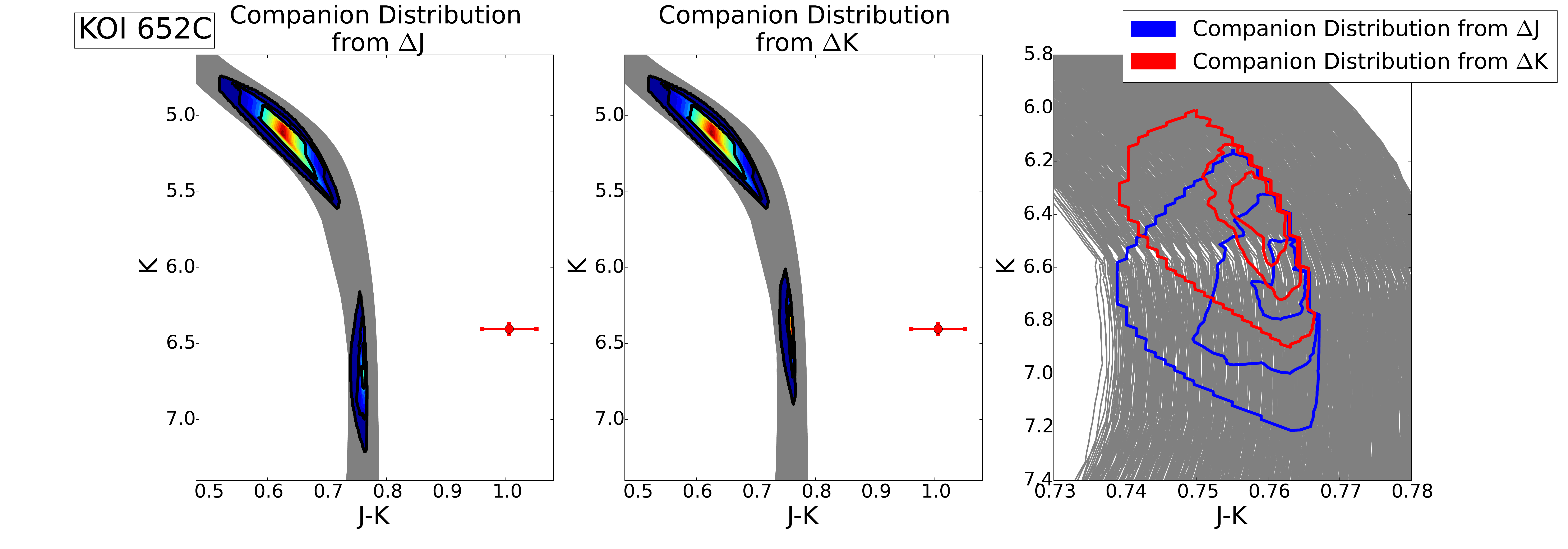}
\caption{\scriptsize Plots showing the results of the imaging data analysis in \S3.1 for
  KOI 652, with two detected companions (B, top two rows; C, bottom
  two rows). Left: Primary KOI
  absolute photometry contours, and companion photometry
  contours, calculated from observed $\Delta$J magnitude and assuming it lies at the same distance and has
  the same age and metallicity as the KOI, mapped on the same (primary
  KOI) isochrone. The red point represents the absolute magnitude and
  ``true'' color for the companion (assuming it is bound), calculated from
  relative color information. The spread in color of the contours
  represents the spread in the normalized probability distribution,
  ranging from 1 (red) to 0 (dark blue). Middle: Same as left, but with companion photometry
  contours calculated from $\Delta$K magnitude. Right: A comparison of
  the overlap between the relative photometry contours of the companion.}
\label{iso_koi652}
\end{figure}

\begin{figure}[h]
\centering
\includegraphics[width=0.95\textwidth]{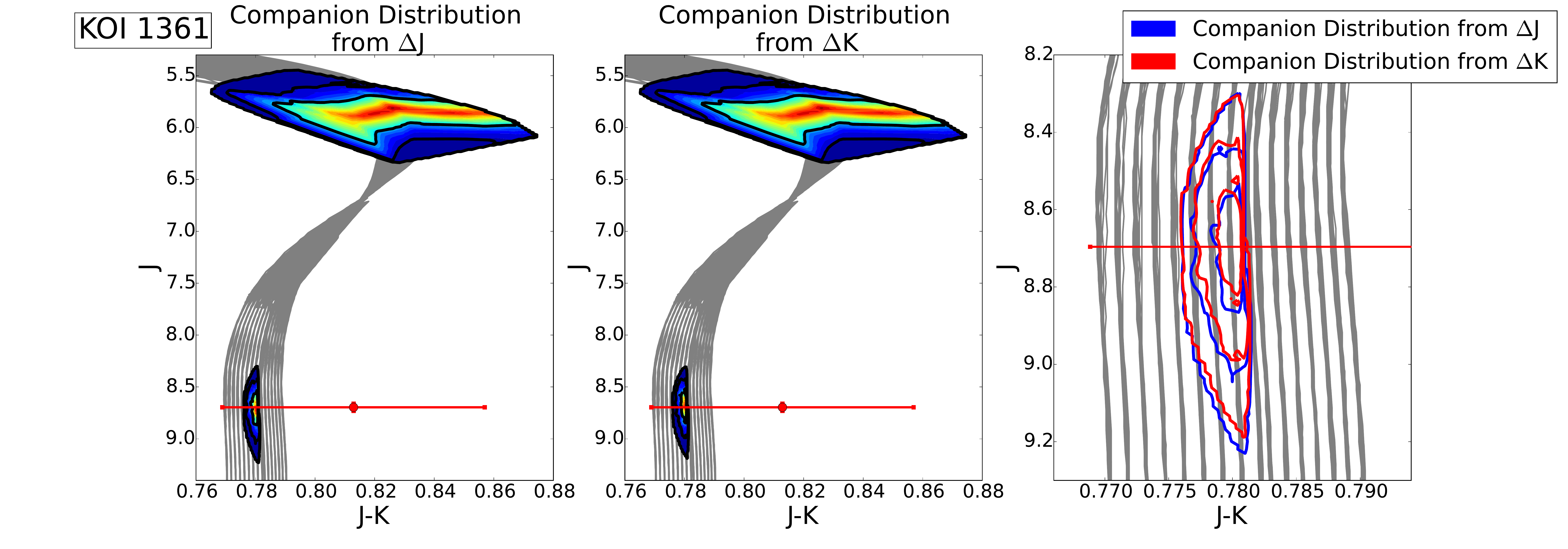}
\quad
\includegraphics[width=0.95\textwidth]{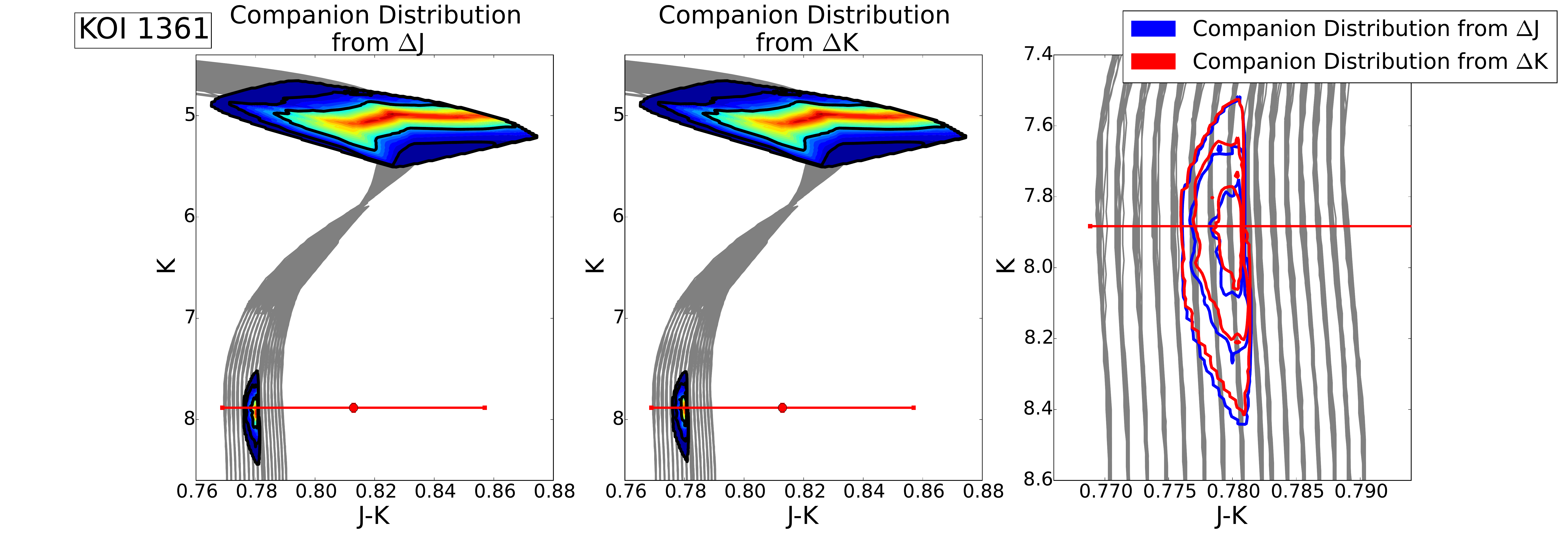}
\caption{Plots showing the results of the imaging data analysis in \S3.1 for
  KOI 1361. Left: Primary KOI
  absolute photometry contours, and companion photometry
  contours, calculated from observed $\Delta$J magnitude and assuming it lies at the same distance and has
  the same age and metallicity as the KOI, mapped on the same (primary
  KOI) isochrone. The red point represents the absolute magnitude and
  ``true'' color for the companion (assuming it is bound), calculated from
  relative color information. The spread in color of the contours
  represents the spread in the normalized probability distribution,
  ranging from 1 (red) to 0 (dark blue). Middle: Same as left, but with companion photometry
  contours calculated from $\Delta$K magnitude. Right: A comparison of
  the overlap between the relative photometry contours of the
  companion. The red point here
is the same as in the left and middle panels.}
\label{iso_koi1361}
\end{figure}

\begin{figure}[h]
\centering
\includegraphics[width=0.95\textwidth]{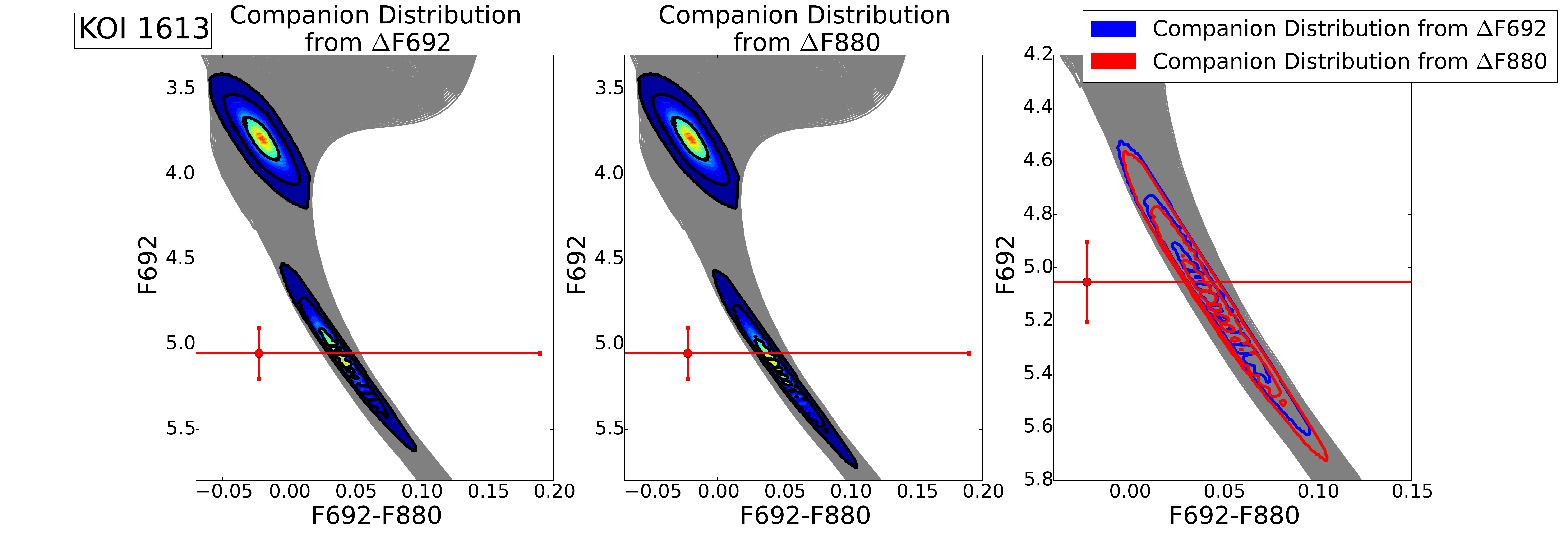}
\quad
\includegraphics[width=0.95\textwidth]{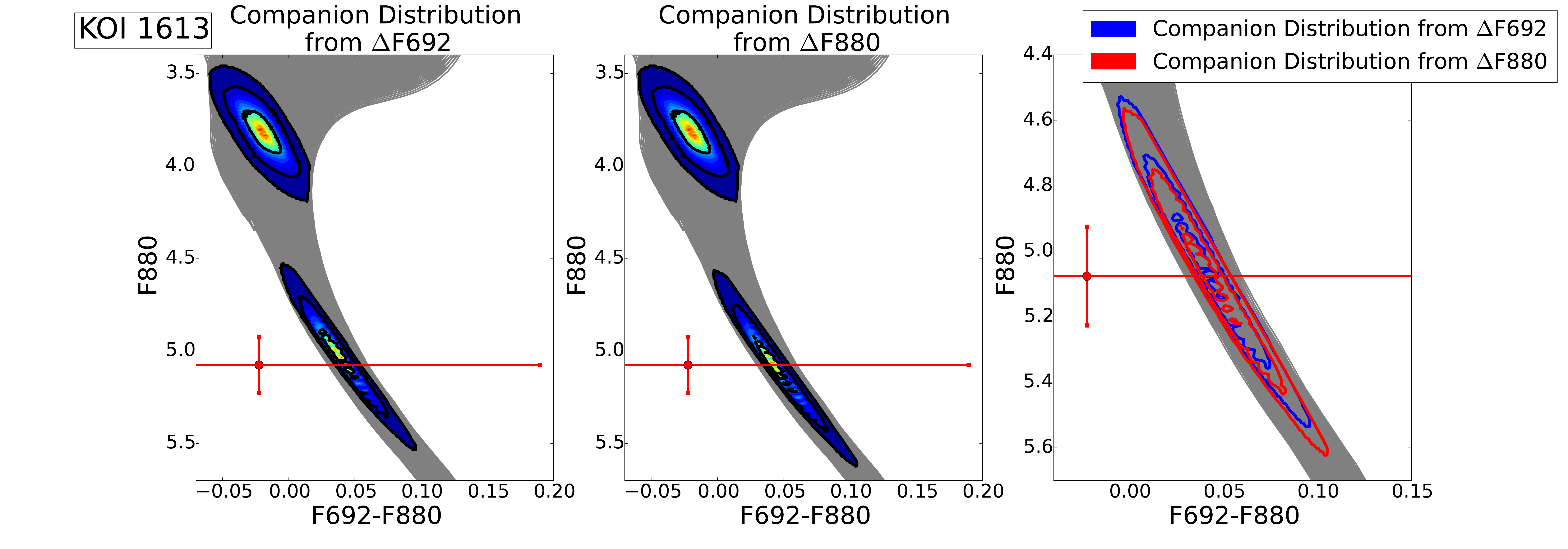}
\quad
\includegraphics[width=0.95\textwidth]{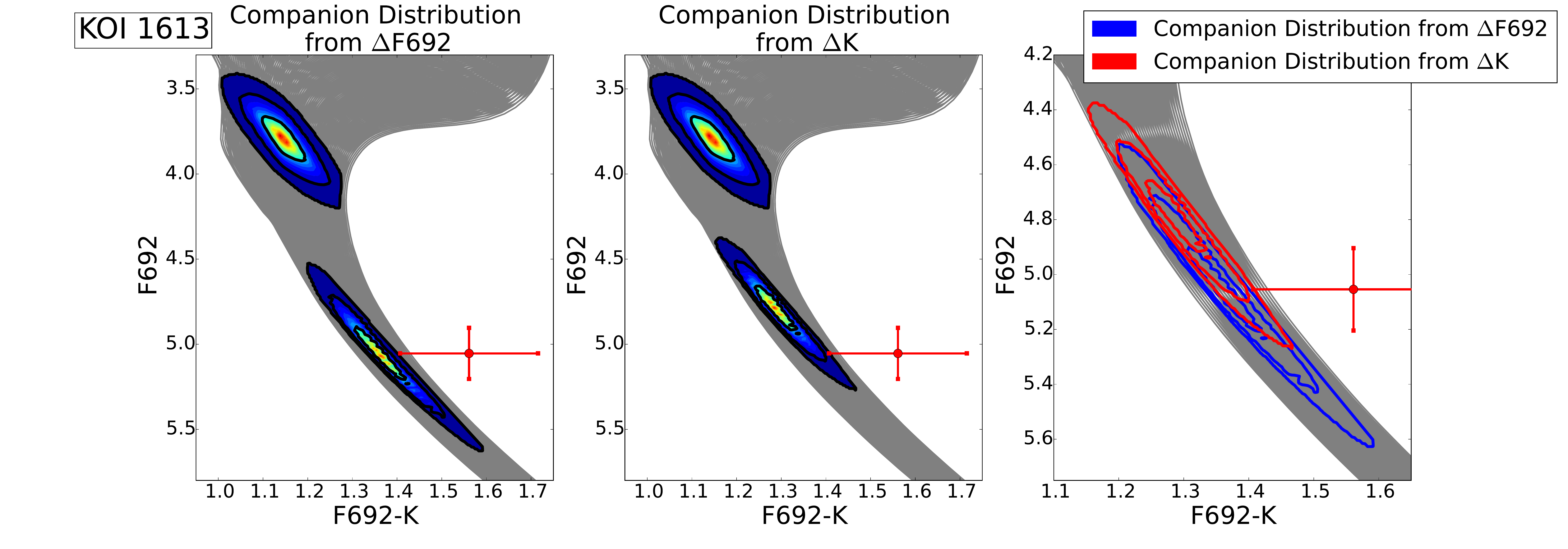}
\quad
\includegraphics[width=0.95\textwidth]{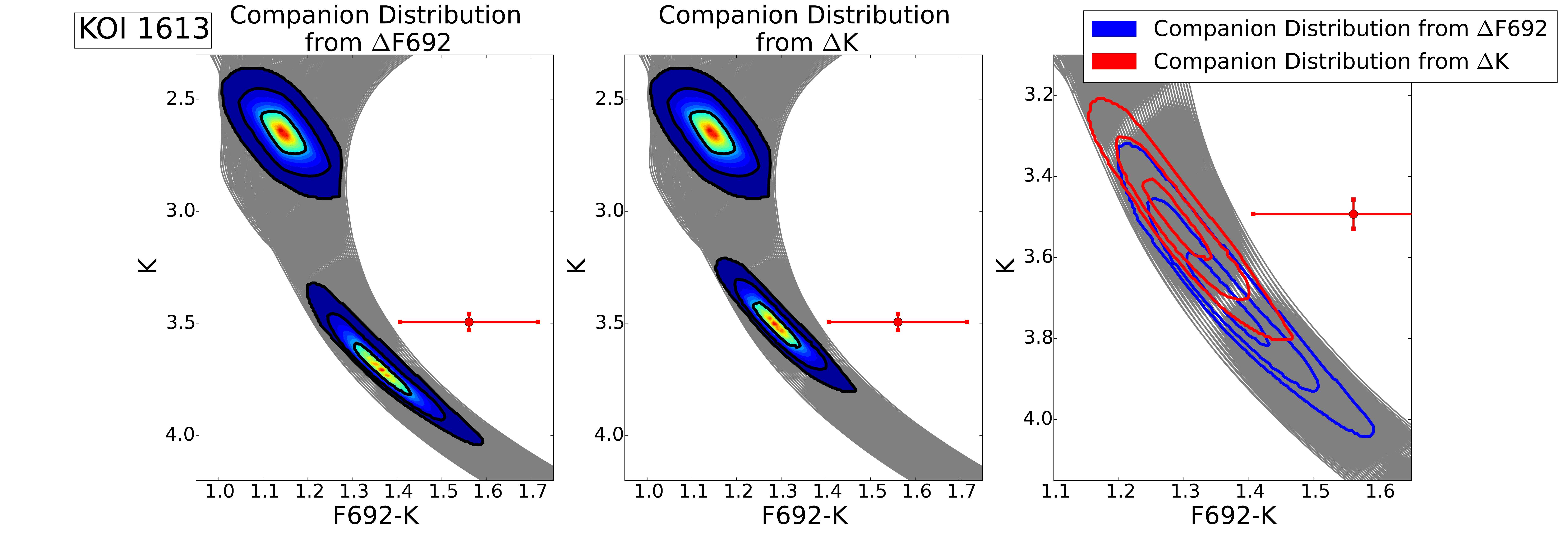}
\caption{\scriptsize Plots showing the results of the imaging data analysis in \S3.1 for
  KOI 1613. Left: Primary KOI
  absolute photometry contours, and companion photometry
  contours, calculated from observed $\Delta$F692 magnitude and assuming it lies at the same distance and has
  the same age and metallicity as the KOI, mapped on the same (primary
  KOI) isochrone. The red point represents the absolute magnitude and
  ``true'' color for the companion (assuming it is bound), calculated from
  relative color information. The spread in color of the contours
  represents the spread in the normalized probability distribution,
  ranging from 1 (red) to 0 (dark blue). Middle: Same as left, but with companion photometry
  contours calculated from $\Delta$F880 (or $\Delta$K, bottom two
  plots) magnitude. Right: A comparison of
  the overlap between the relative photometry contours of the
  companion. The red point here
is the same as in the left and middle panels.}
\label{iso_koi1613}
\end{figure}

\begin{figure}[h]
\centering
\includegraphics[width=0.95\textwidth]{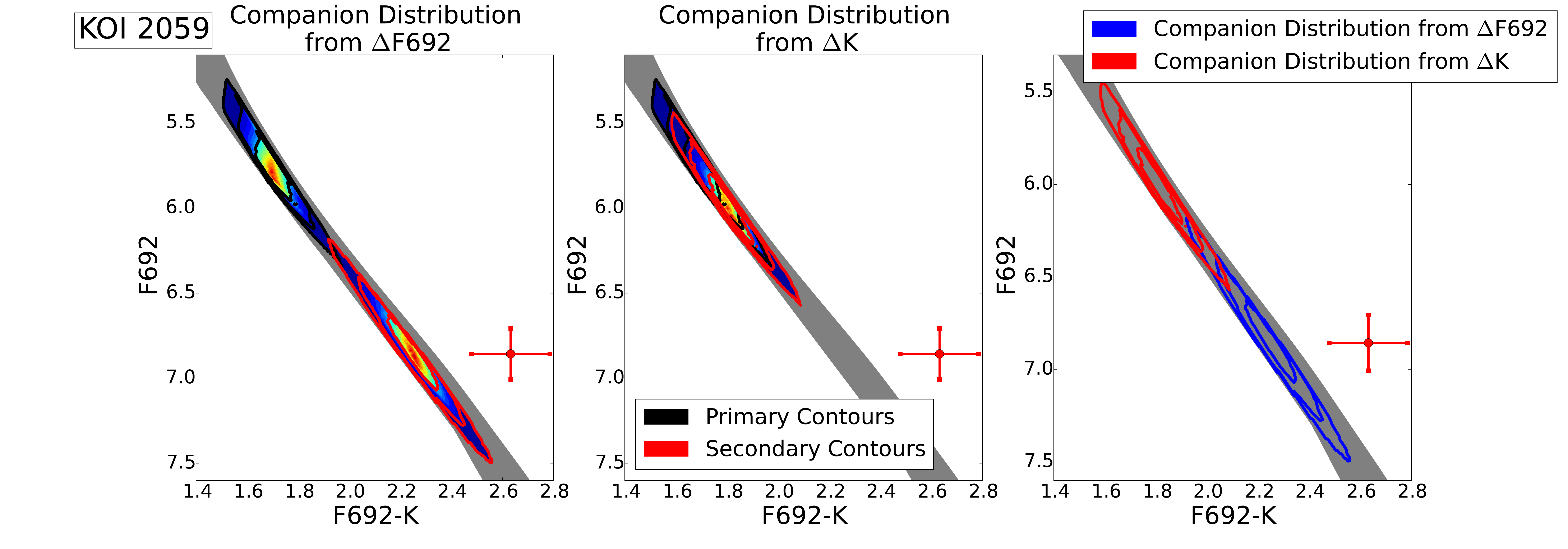}
\quad
\includegraphics[width=0.95\textwidth]{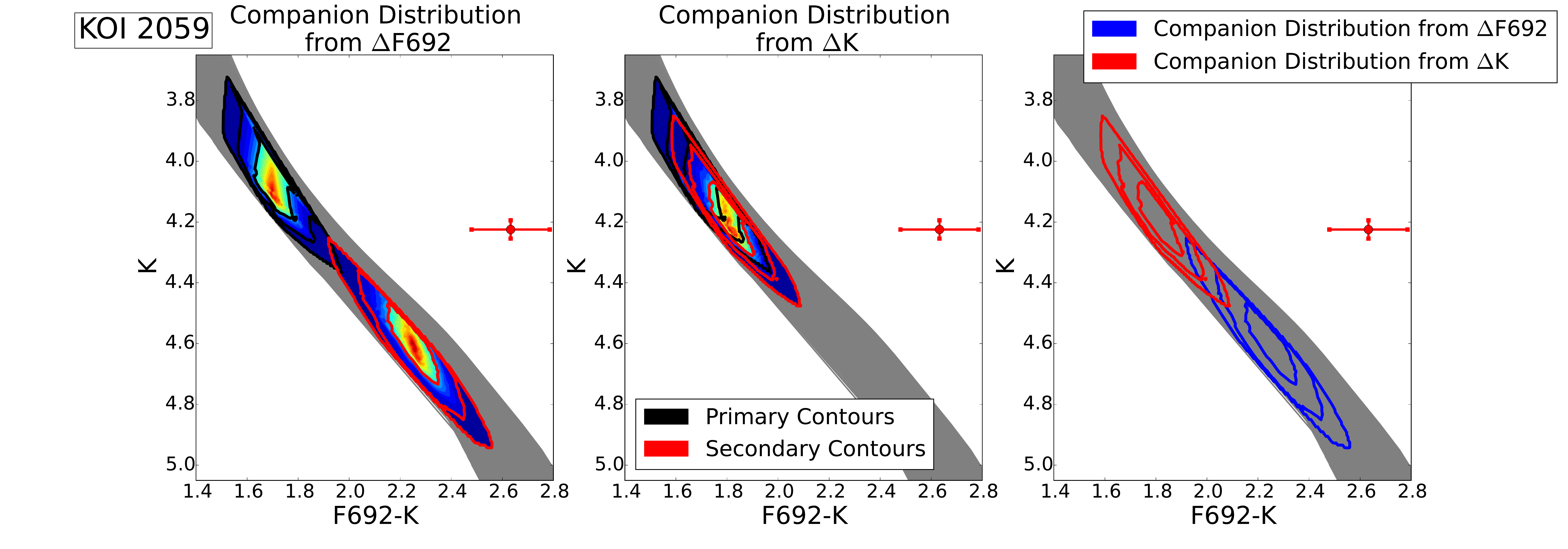}
\caption{ Plots showing the results of the imaging data analysis in \S3.1 for
  KOI 2059. Left: Primary KOI
  absolute photometry contours, and companion photometry
  contours, calculated from observed $\Delta$F692 magnitude and assuming it lies at the same distance and has
  the same age and metallicity as the KOI, mapped on the same (primary
  KOI) isochrone. The red point represents the absolute magnitude and
  ``true'' color for the companion (assuming it is bound), calculated from
  relative color information. The spread in color of the contours
  represents the spread in the normalized probability distribution,
  ranging from 1 (red) to 0 (dark blue). Middle: Same as left, but with companion photometry
  contours calculated from $\Delta$K magnitude. Right: A comparison of
  the overlap between the relative photometry contours of the
  companion. The red point here
is the same as in the left and middle panels.}
\label{iso_koi2059}
\end{figure}

\begin{figure}[h]
\centering
\includegraphics[width=0.95\textwidth]{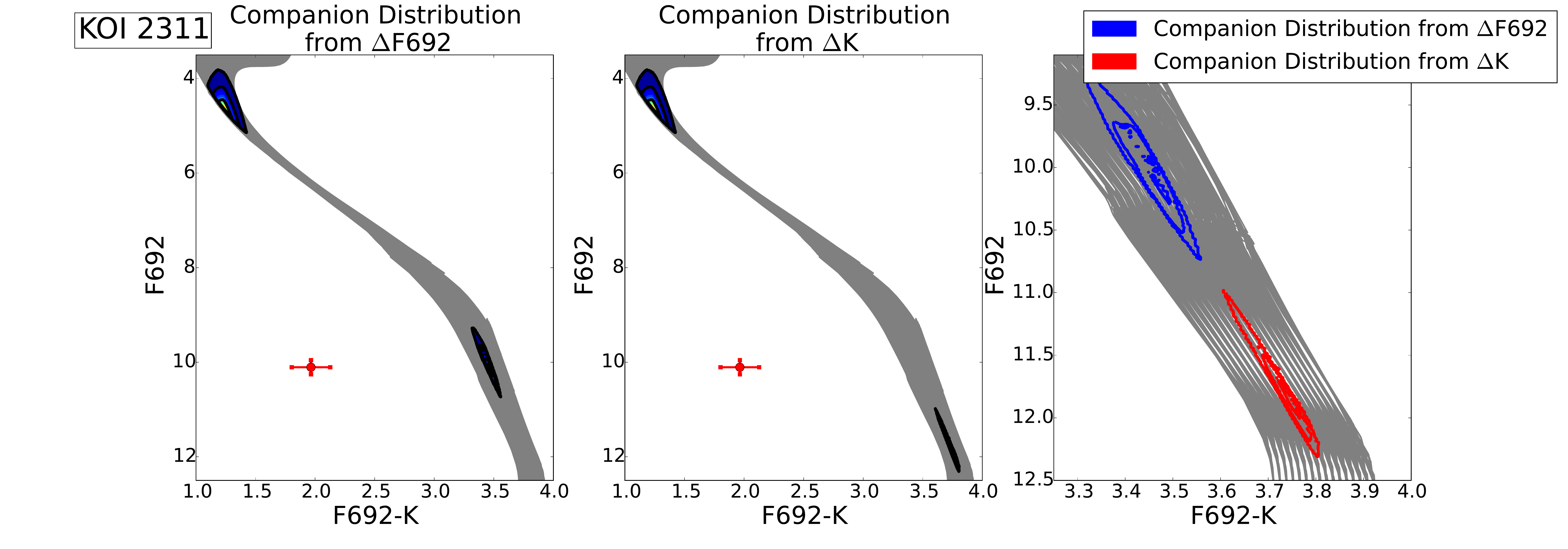}
\quad
\includegraphics[width=0.95\textwidth]{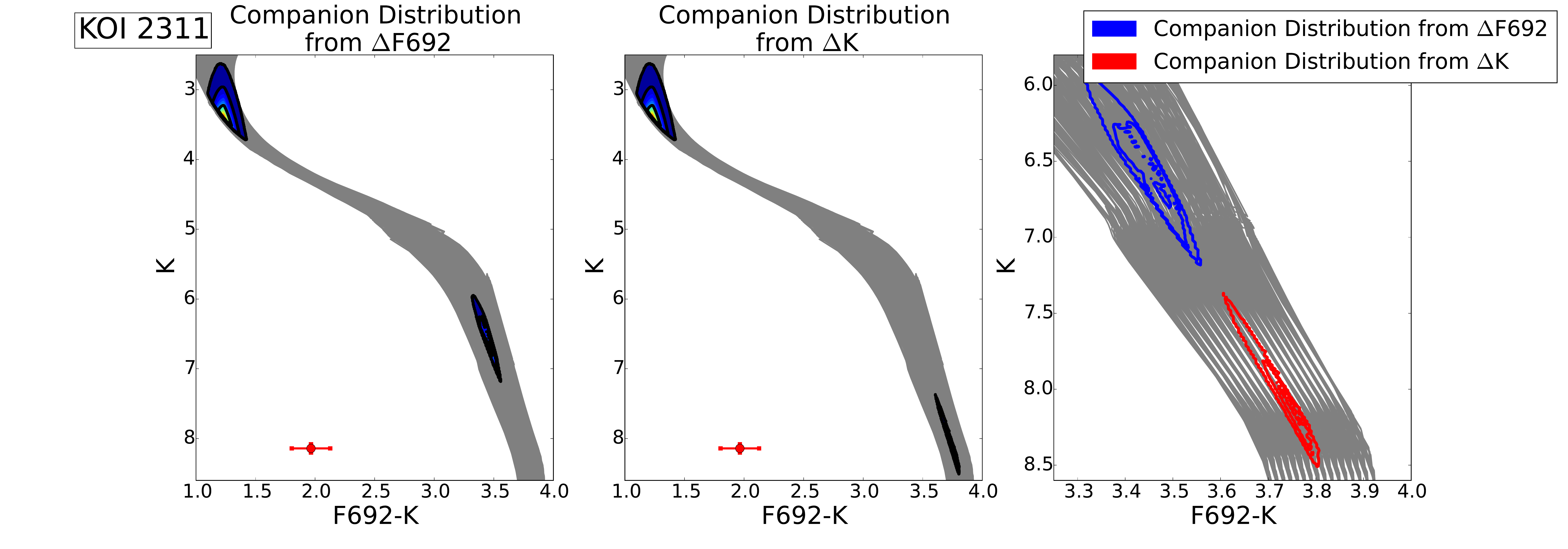}
\quad
\includegraphics[width=0.95\textwidth]{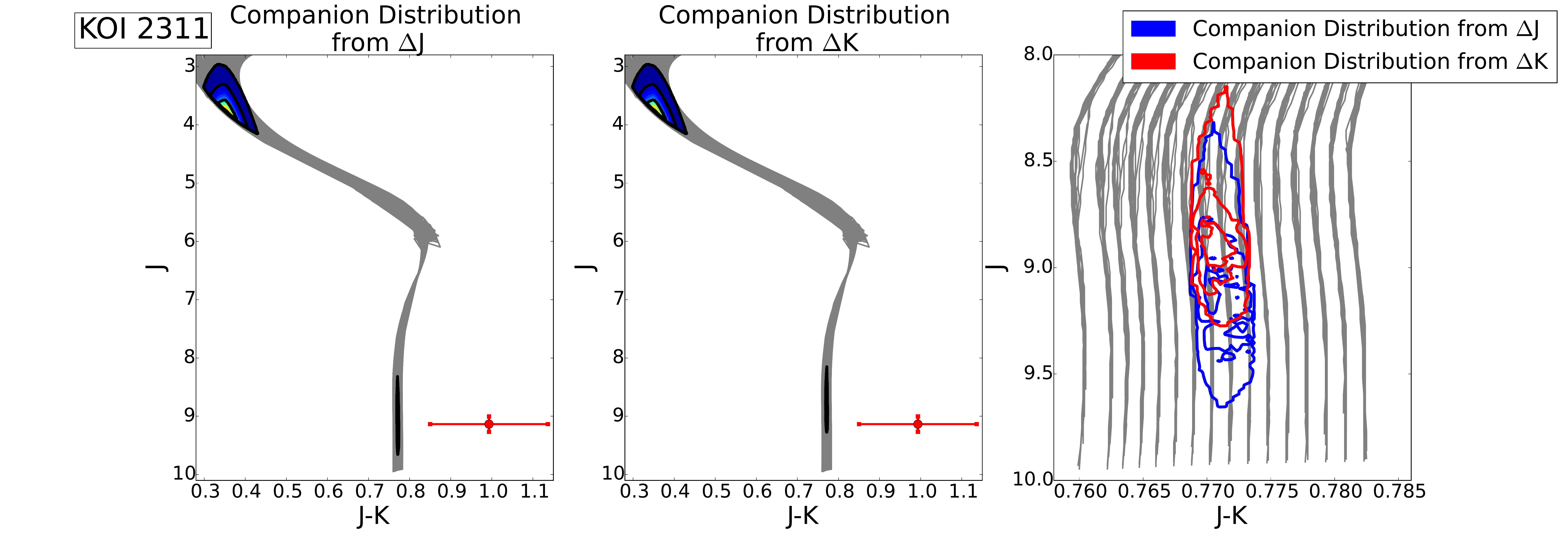}
\quad
\includegraphics[width=0.95\textwidth]{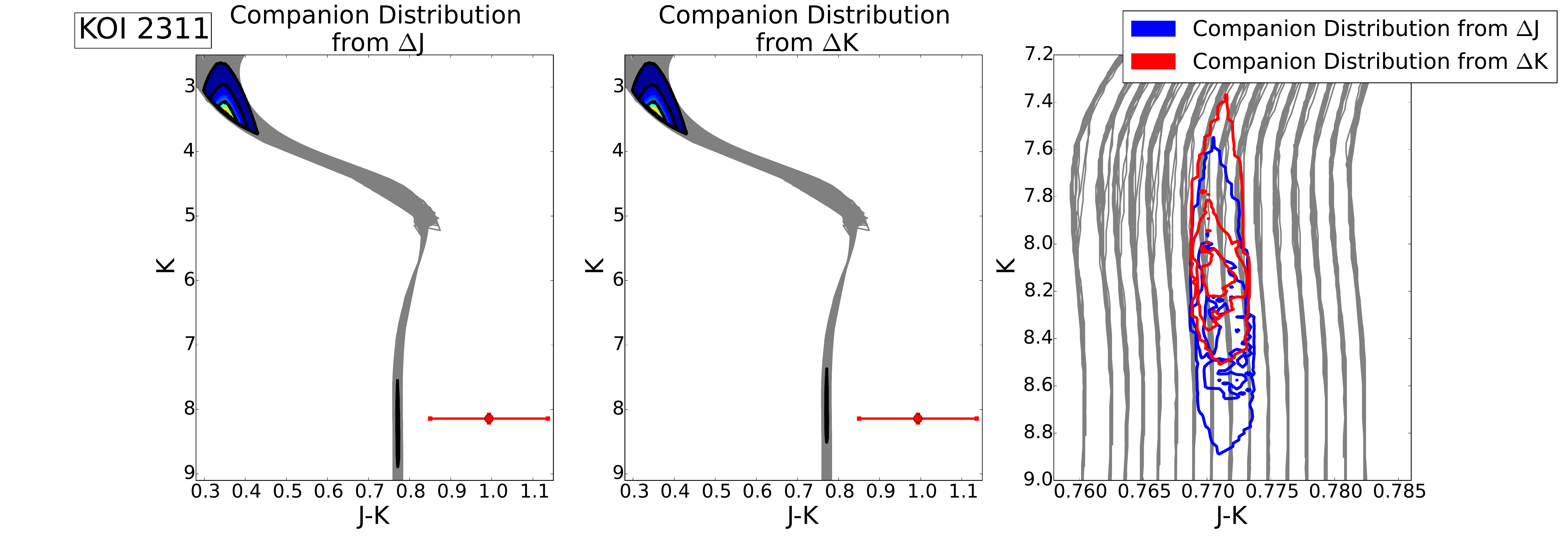}
\caption{\scriptsize Plots showing the results of the imaging data analysis in \S3.1 for
  KOI 2311. Left: Primary KOI
  absolute photometry contours, and companion photometry
  contours, calculated from observed $\Delta$F692 (or $\Delta$J, bottom
  two plots) magnitude and assuming it lies at the same distance and has
  the same age and metallicity as the KOI, mapped on the same (primary
  KOI) isochrone. The red point represents the absolute magnitude and
 `` true'' color for the companion (assuming it is bound), calculated from
  relative color information. The spread in color of the contours
  represents the spread in the normalized probability distribution,
  ranging from 1 (red) to 0 (dark blue). Middle: Same as left, but with companion photometry
  contours calculated from $\Delta$K magnitude. Right: A comparison of
  the overlap between the relative photometry contours of the
  companion.}
\label{iso_koi2311}
\end{figure}

\begin{figure}[h]
\centering
\includegraphics[width=0.95\textwidth]{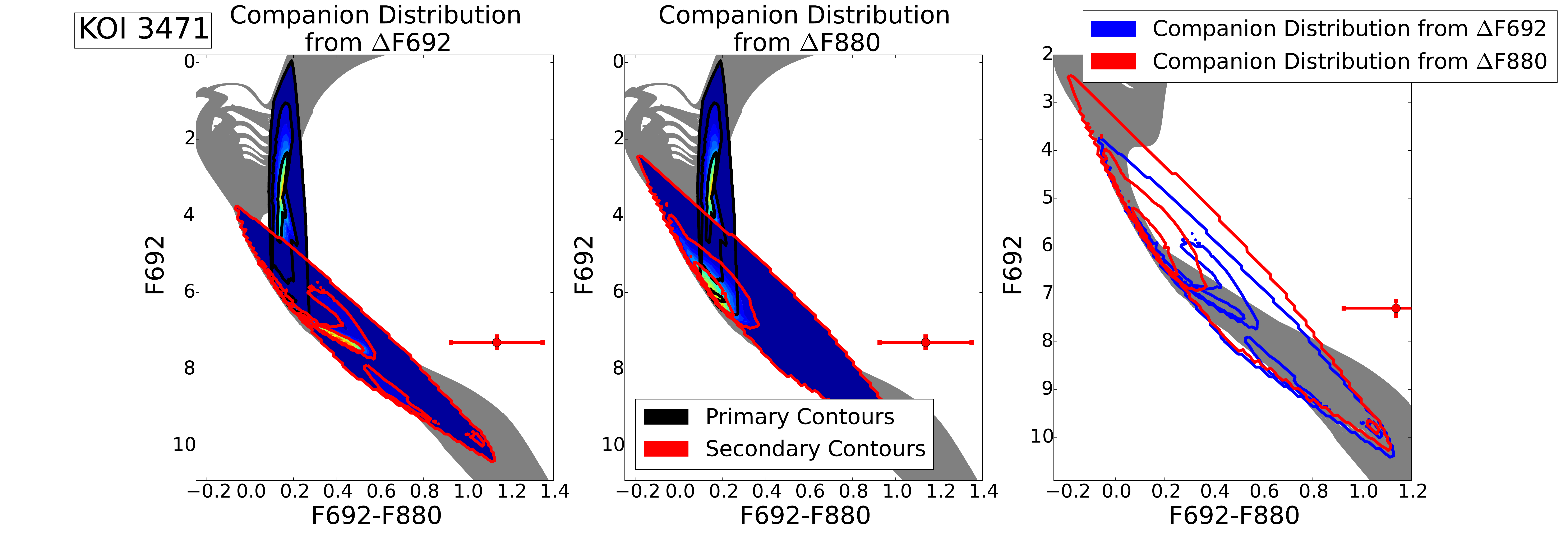}
\quad
\includegraphics[width=0.95\textwidth]{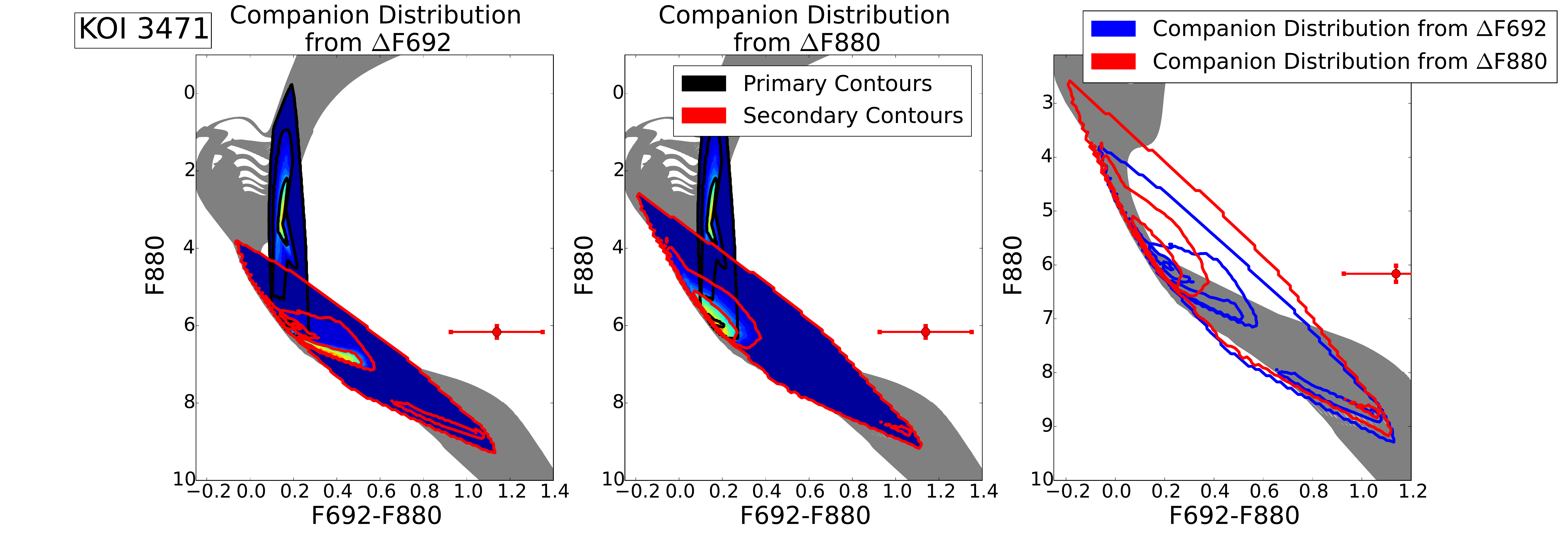}
\quad
\includegraphics[width=0.95\textwidth]{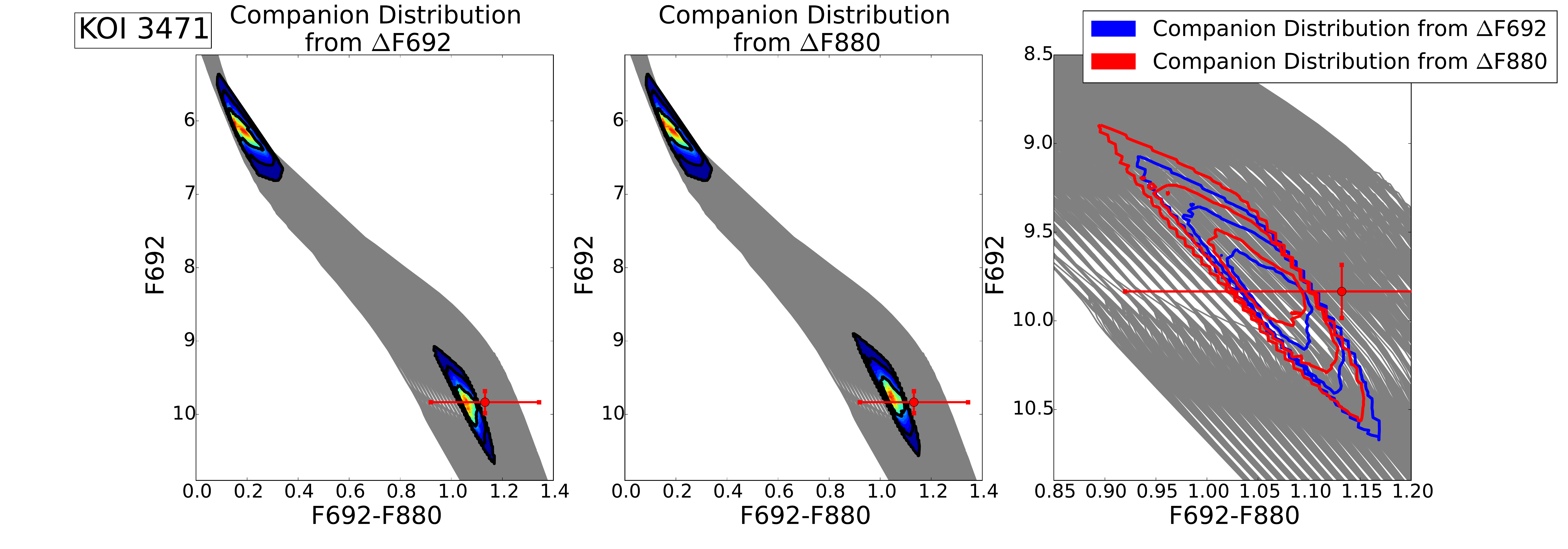}
\quad
\includegraphics[width=0.95\textwidth]{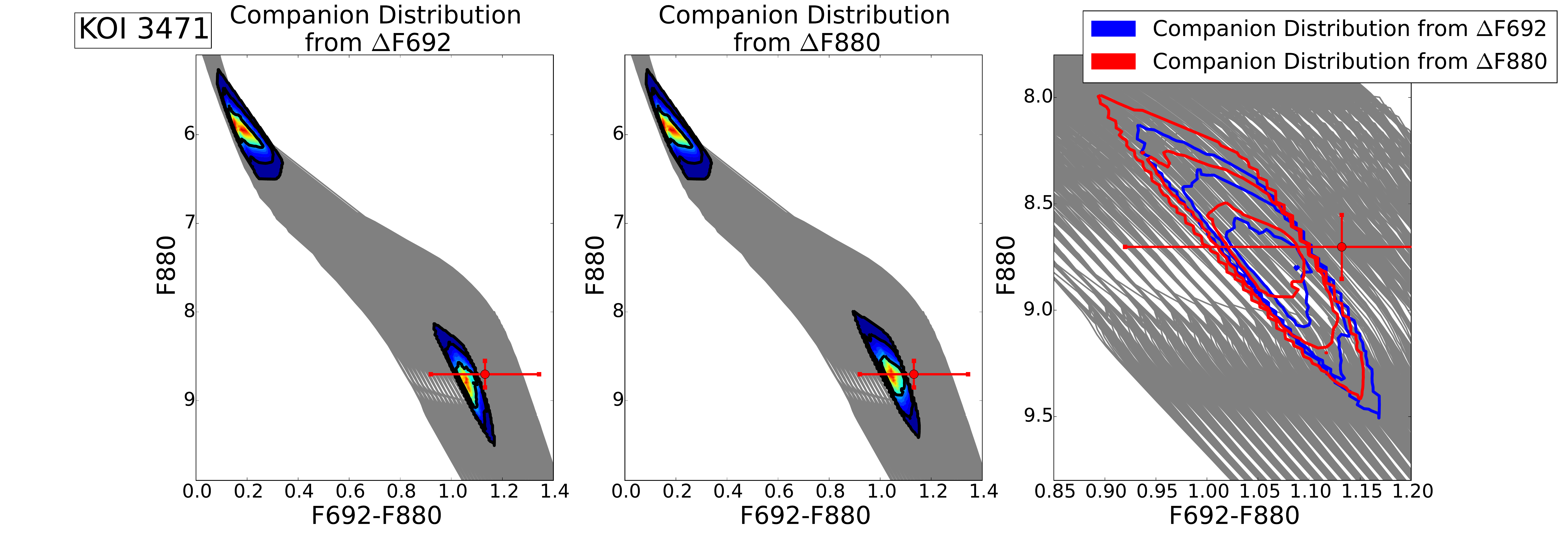}
\caption{\scriptsize Plots showing the results of the imaging data analysis in \S3.1 for
  KOI 3471. The top two rows show results based on the Huber et al. (2014)
  parameters for KOI 3471, classifying it as a subgiant. The bottom
  two rows show results based on a dwarf-like log $g$ for KOI 3471
  (4.6$\pm$0.02); see \S4.2.2. Left plots: Primary KOI
  absolute photometry contours, and companion photometry
  contours, calculated from observed $\Delta$F692 magnitude and assuming it lies at the same distance and has
  the same age and metallicity as the KOI, mapped on the same (primary
  KOI) isochrone. The red point represents the absolute magnitude and
  ``true'' color for the companion (assuming it is bound), calculated from
  relative color information. The spread in color of the contours
  represents the spread in the normalized probability distribution,
  ranging from 1 (red) to 0 (dark blue). Middle plots: Same as left, but with companion photometry
  contours calculated from $\Delta$F880 magnitude. Right plots: A comparison of
  the overlap between the relative photometry contours of the
  companion. The red point here
is the same as in the left and middle panels.}
\label{iso_koi3471}
\end{figure}

For stars with multi-band photometry, we plot in Figures \ref{iso_koi5}-\ref{iso_koi3471} the primary and
companion probability distributions 
in color-magnitude space based on the absolute magnitudes associated
with each mass point and isochrone. To plot these distributions, we
perform a linear interpolation of the isochrone mass points onto a
regular grid of color–-magnitude--metallicity, then plot a slice in
metallicity at the input primary [Fe/H]. We overplot a set of
isochrones within $\pm 1\sigma$ in metallicity of the primary. We also
plot (in red) the ``true'' photometric points for the companion stars,
based on the differential photometry and the primary model absolute
magnitudes in each filter. Assuming the companions are bound, their
extinction corrections will be the same as those of their primary KOIs.

The multi-color photometry analysis suggests that the companions to
KOIs 5, 1361, and 3471 are likely to be bound, the companion to KOI 1613 may
be bound, and the companions to KOI 652, 2059, and 2311 are
unlikely to be bound. 
Thus the derived companion effective temperatures listed in Table 2
may only be valid for KOIs 5, 1361, 1613, and 3471 while the temperatures
calculated for the companions to KOIs 652, 2059, and 2311 are
unlikely to be accurate. In \S4.2 we consider how the bound versus
unbound nature of these companions relates to the detections reported
by K15.

\section{Discussion}

\subsection{Expected Overlap Between Techniques}

Before examining the measured results of the companions detected by
spectroscopy versus imaging, it is worth examining the theoretical
detectability of any/all binary stars in the list of KOIs. We extend
the binary star simulations published in Horch et al. (2014) to
include predictions for the radial velocities seen in each binary. The
Horch et al. simulation constructs a set of binary stars with properties representative of those expectated among the
{\it Kepler} exoplanet targets. Primary stars are modeled using the
TRILEGAL Galaxy model (Girardi et al. 2005), but including only those
stars within a restricted log($g$) range in order to simulate the
pre-selection of dwarfs that dominate the \textit{Kepler} target list. Secondaries are assigned random masses based on the binary
mass distribution found by Raghavan et al. (2010) and orbital periods
and eccentricities are assigned randomly to satisfy the distributions
found in Duquennoy \& Mayor (1991).  The remaining orbital elements (the cosine of the
inclination, angle of the ascending node, the angle between the line
of nodes and the semi-major axis, the epoch of the observation and the
time of periastron passage) were assigned using a uniform random
distribution over all possible values.  Observed quantities are
predicted for each binary, including the angular separations on the
sky, magnitudes and the $\Delta$RV. Note that the simulated RVs are calculated for
a random point in the orbit, so range from zero to the maximum RV that
would be measured over the entire binary orbit, to best match what is
measured from a single spectrum taken in a population of binaries. 

\begin{figure}[h]
\centering
\includegraphics[width=1.0\textwidth]{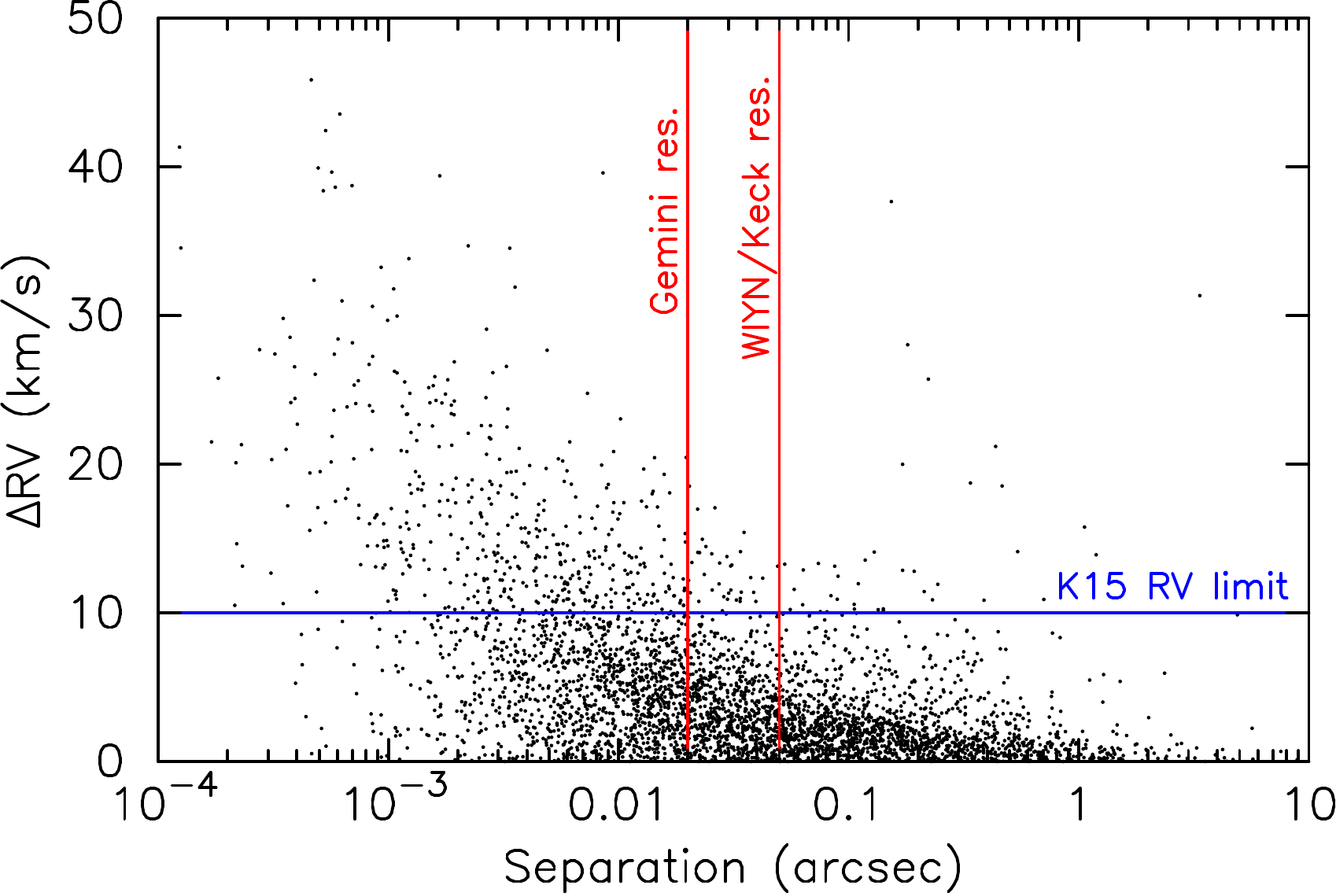}
\caption{Radial velocity difference between components in a set of 7958
simulated {\it Kepler} binary stars is plotted vs. their angular
separation.  The two red lines indicate the resolution limits expected
for optical speckle imaging at the \textit{Gemini}~8m telescope (0.02$\arcsec$)
and optical speckle imaging at the \textit{WIYN}~3.5m or near-infrared adaptive
optics imaging at the \textit{Keck}~10m telescopes ($\sim0.05\arcsec$).
Binaries with angular separations exceeding these lower limits can be
spatially resolved.  The blue line represents the 10~km~s$^{-1}$ lower
limit to the difference in radial velocity between binary components
for spectroscopic detection using the methods of Kolbl et al. (2015). The figure shows that the two complementary techniques should detect
largely separate populations of binaries and that only a small
fraction of the total ($\sim0.5$\%) can be detected simultaneously
using both techniques.}
\label{RVvsSep}
\end{figure}

Figure~\ref{RVvsSep} shows the results of the simulation of 7958 {\it Kepler}
binaries.  The radial velocity difference between components of each
binary are plotted vs. their angular separation.  Vertical red lines
show the resolution limits for different imaging data sets.  Secondary
stars in binaries whose separations lie between the resolution limits
and the edge of the instrument's field-of-view should be detected,
except in cases of very large magnitude differences.  The horizontal
blue line shows the nominal 10~km~s$^{-1}$ radial velocity difference
needed to detect secondaries using the K15 methods.  Secondary stars
in binaries with relative radial velocities exceeding this threshold
and within $0.5-20\%$ of the flux of the primary star are expected to
be detected.  As discussed earlier, some M dwarf companions to hotter
stars may also be detectable, even with lower relative velocities.
Figure~\ref{RVvsSep} shows that the imaging and spectroscopy methods should detect
quite different secondary populations.  

The effectiveness of different
techniques can be quantified based on the model. For
example, the fraction of secondaries whose separations are resolvable
is 59\% for speckle imaging at \textit{Gemini}.  However, the faintest
stars fall below the detection limits and a few binary pairs have
large enough separations to fall outside the imaging field, lowering
the percentage of all secondaries that would be detcted by {\it
  Gemini} to about 32\%. 
The K15 technique
should mainly detect secondaries whose angular separations would be
too small for imaging surveys. At such separations, both components
would be expected to fall within the spectrograph slit. The fraction of all binaries with
$\Delta$RV$>10$~km~s$^{-1}$ is 6.7\%.  The fraction of these
recovered by K15 is expected to be lowered to 5$-$5.5\% by
requiring that the secondary flux be at least $0.5-1$\% of the
primary (K15's minimum detectable flux ratio).  The binary parameter space in which both K15 and imaging
surveys are expected to detect the same secondaries lies in the upper
right hand part of Figure~\ref{RVvsSep}, which is sparsely populated by the
simulation. The fraction of secondaries simultaneously recoverable
using both techniques is predicted by this simulation to be a mere $\sim 0.5$\%.

However, any binaries in the sample of 11 stars considered in this
work, having both K15 and imaging detected companions, are more likely than a random binary KOI to
be within that subsample of $\sim 0.5$\%. In K15's sample of 1160 KOI
spectra, they find 63 doubles, 5.4\% of their sample. 
This fraction is in agreement with the number of recoverable binaries
predicted in our simulation if the binary fraction among KOI stars is $\sim$50\% and if the doubles detected by K15 are
composed of comparable numbers of binary and co-aligned field stars.

\subsection{Comparison of Spectroscopy vs. Imaging Samples}

\subsubsection{Considering the Overall Sample}
With companion temperatures and flux ratios derived from spectroscopy (K15) and
imaging (this work), we can try to directly compare the measured results of the
two techniques. In Table \ref{tab2} we list the K15 results for the
overlapping sample -- KOIs with companions detected by
K15 that also have companions detected in imaging data -- as well as the results of our
analysis of the KOI companions (\S3.1). 

\begin{figure}[ht!]
\centering
 \subfigure{\includegraphics[width=0.5\textwidth]{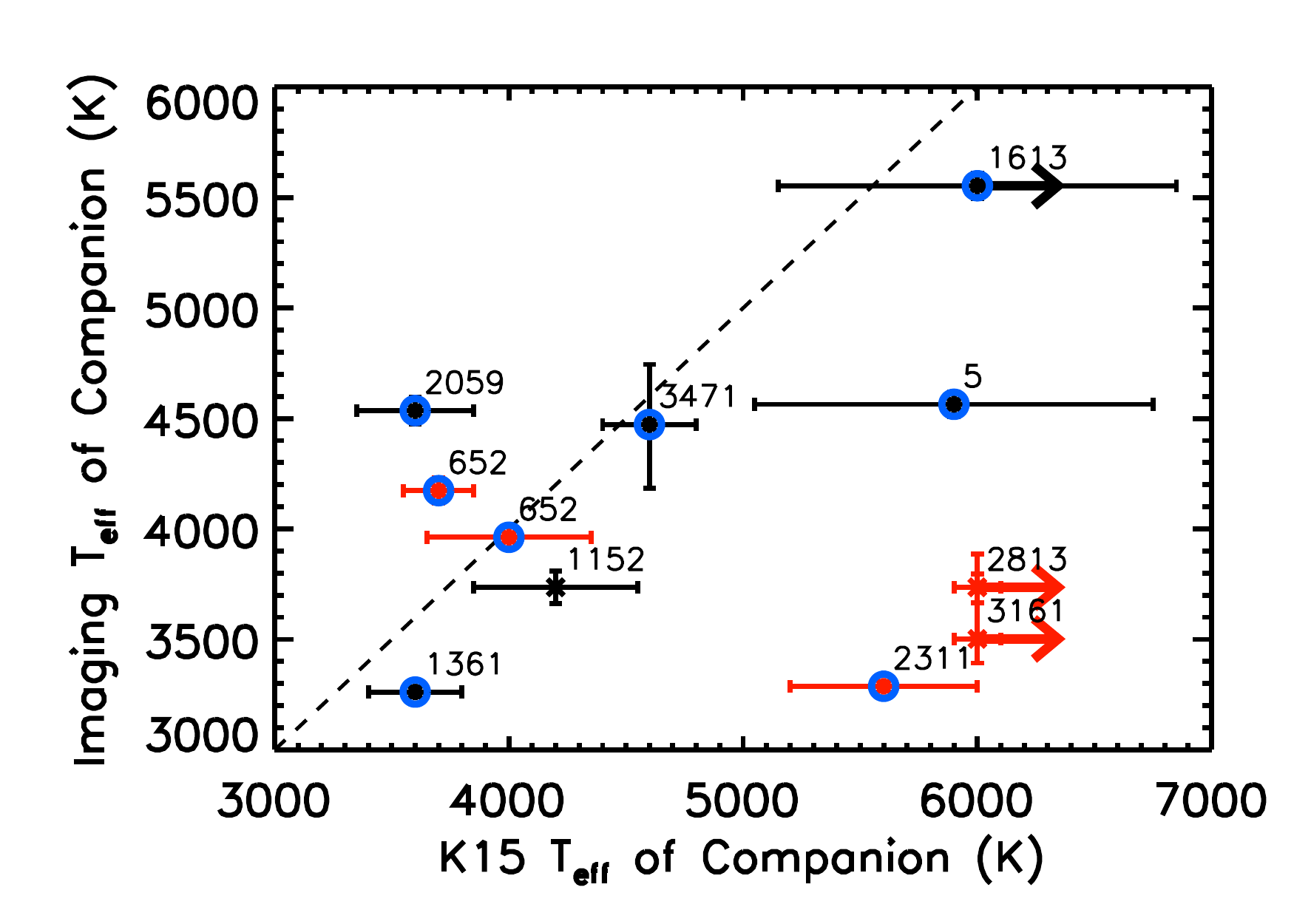}}
\hspace{-30pt}
\subfigure{\includegraphics[width=0.5\textwidth]{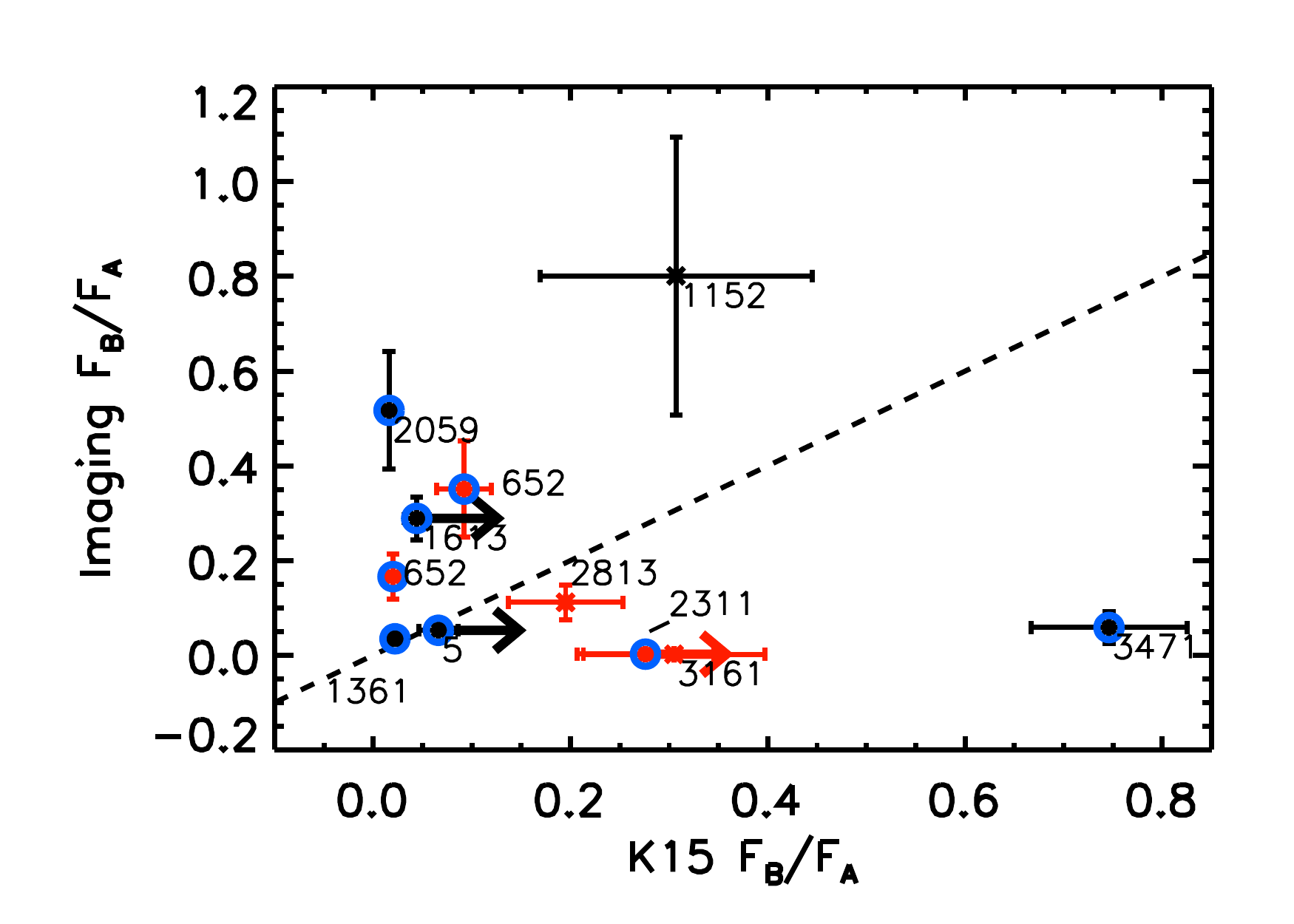}}
\hspace{-30pt}
 \subfigure{\includegraphics[width=0.5\textwidth]{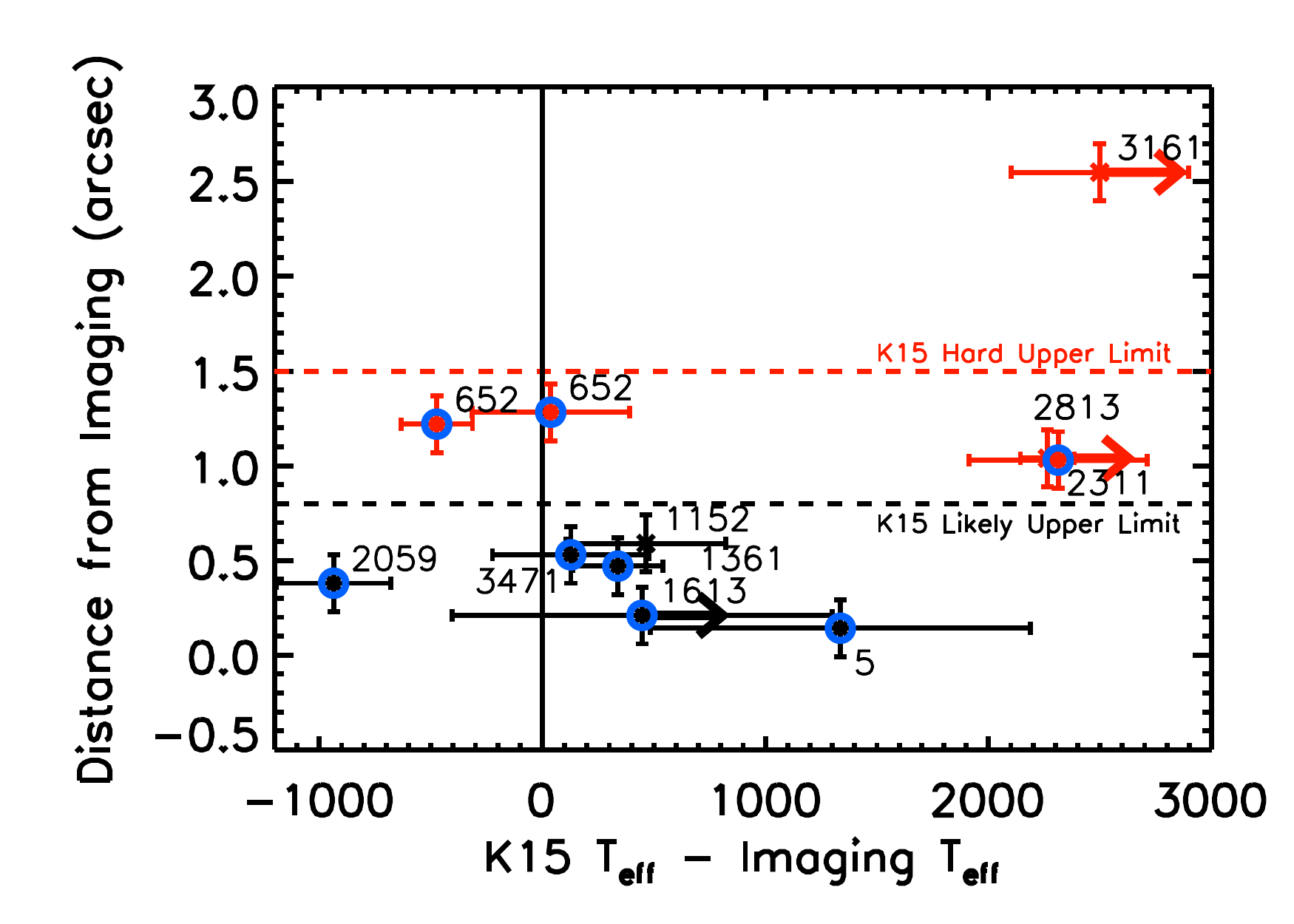}}
\hspace{-30pt}
\subfigure{\includegraphics[width=0.5\textwidth]{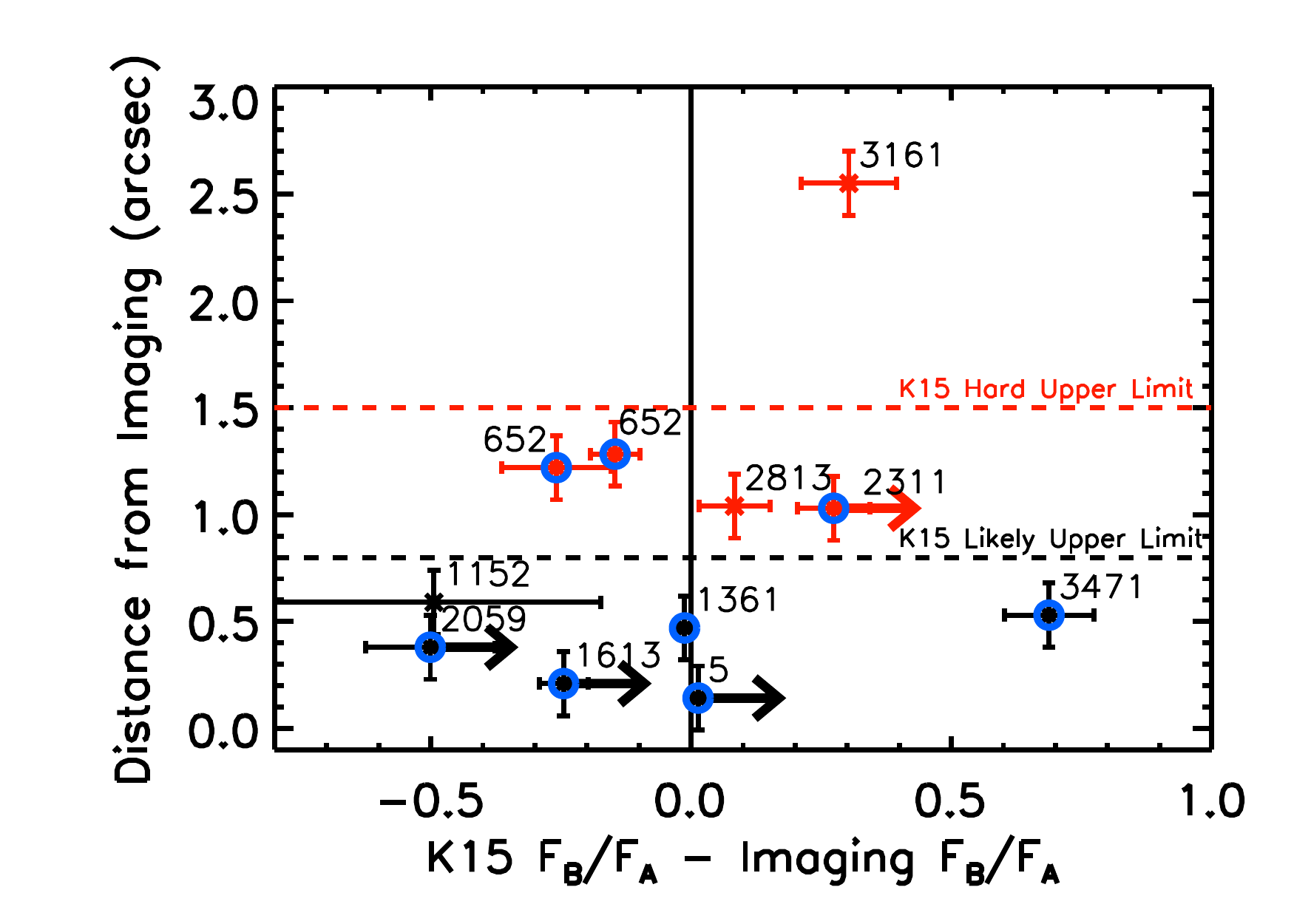}}
\caption{Plots showing parameter comparisons for companions
 found through K15's analysis, and derived from our
 analysis of imaging data. The labels refer to the companion's primary
 KOI. Blue circles indicate KOIs with imaging data in multiple bands,
 and red symbols indicate the imaging-detected companion is $>$0.8$\arcsec$
 away from the primary KOI. The values plotted here for the companion to
 KOI 3471 assume a subigant primary star. \textit{Note that the ``imaging'' values are those
 derived from our analysis in \S3.1, and thus may be incorrect if the
 companion is unbound.} \textit{Top left}: T$_{eff}$ values of the companions. \textit{Top right}: Flux ratios
 (companion/primary) of the companions. The flux ratios measured from imaging data are in the $Kepler$
 bandpass. \textit{Bottom left}: Difference in derived T$_{eff}$ values, versus the separation as measured
  from imaging data (averaged over all detections). Dashed horizontal lines designate the separation limits
reported in K15. \textit{Bottom
  right}: Difference in derived flux ratios
  of companions, versus the separation as measured
  from imaging data (averaged over all detections). Dashed horizontal lines designate the separation limits
reported in K15. }
\label{secondary_comp}
\end{figure}

First, we assess the overlapping sample for similarities and
differences in their derived parameters.
Figure
\ref{secondary_comp}, top row, shows the T$_{eff}$ values and flux ratios of
the companions detected using one method versus the other. The K15
flux ratios are measured across the HIRES wavelength range (4977-7990
\AA), avoiding regions with telluric pollution and the interstellar
sodium D lines; the imaging flux
ratios are reported for the $K_p$ bandpass, $\sim$4000-9000 \AA. A dashed line designates
slope$=$1, blue circled points indicate multiple bands
(filters) of imaging data (e.g., $K$ and 692 nm), and red points
indicate separations $>$0.8$\arcsec$ for the imaging-detected companion. There is some
agreement between the companion T$_{eff}$ values derived from
different methods, especially considering the large K15 errors on the
companions to KOIs 5 and 1613. The companion to KOI 2059 and one of
the companions to KOI 652 detected by K15 are cooler than the
companions detected by imaging (according to the temperatures derived
in \S3.1), while the companions to KOI 2311, 2813, and 3161 detected
by K15 are much hotter; the companions to KOI 2813 and 3161 detected
by K15 actually have only lower limits to their derived
T$_{eff}$. There is less agreement between the companion flux ratios derived
from different methods (right, top plot); only two, possibly three companions to KOIs
are consistent with the slope$=$1 line. The flux ratios of the
imaging-detected companions to KOIs 652,
1152, 1613, and 2059 are higher (via the analysis in this
work) versus the K15 detections and analysis, while the
imaging-detected companions to KOIs 2311,
2813, 3161, and 3471 have lower flux ratios versus the spectroscopy-detected
companions reported in K15. 

To try to understand the physical explanation for the disagreements between the two analysis methods, we plot in Figure \ref{secondary_comp}, bottom row, the
differences between the companion T$_{eff}$ values and flux ratios derived from the
different methods versus the primary-to-companion distance, measured from the
imaging data (averaged across wavebands). A vertical solid line marks
zero difference, and dashed horizontal lines mark the likely (0.8$\arcsec$)
and hard (1.5$\arcsec$) upper limits on detectable separation from
K15. Again, the blue circled points indicate a companion with multiple bands
(colors) of imaging data, and red points indicate that the
imaging-detected companion has a separation $>$0.8$\arcsec$. Based on the hard upper limits for companion
separation from K15, we might expect the
parameters of the companion detected around KOI 3161 by K15 versus
from imaging data to differ --
at such a large separation light from either primary or companion
stars may not be fully in the \textit{Keck I/HIRES} slit. Similarly, we might
expect the parameters of companions to KOI 652, 2311, and 2813 to
differ, since they are farther than the ideal separation
limit of 0.8$\arcsec$ from K15. However, while the companions to KOI 2311 and
2813 detected by K15 versus the companions detected by imaging clearly differ in both T$_{eff}$ and flux ratio (see also top panel of
this figure), the companions to KOI 652 detected by either method have similar derived
T$_{eff}$ values. There is some abiguity about companions to KOI 652
because K15 detects three companions (and a fourth, which was too low
S/N to be officially reported in K15), whereas imaging detects
two. K15 reports their third detected companion to KOI 652 as cooler
and fainter (lower flux ratio) than the other two companions they
detect, so replacing one of the other companions'
parameters with the cooler/dimmer parameters would only increase
the contrast between K15-derived parameters and the results from
the imaging analysis in \S3.1. Furthermore, companions
to KOIs detected by imaging that are well within the 0.8$\arcsec$ separation limit still have
discrepant parameters between the two observation and analysis methods (all black
points in bottom two panels). 

In summary, from Figure \ref{secondary_comp} there are significant
differences between the companion parameters derived from the spectroscopic and
imaging analyses, yet these differences do not show an obvious pattern
(e.g., companions detected in one method with derived cooler
temperatures also have lower derived flux ratios), 
or dependence on distance from the primary KOI.

\subsubsection{Considering KOIs Individually}
\begin{figure}[h]
\centering
\subfigure{\includegraphics[width=0.53\textwidth]{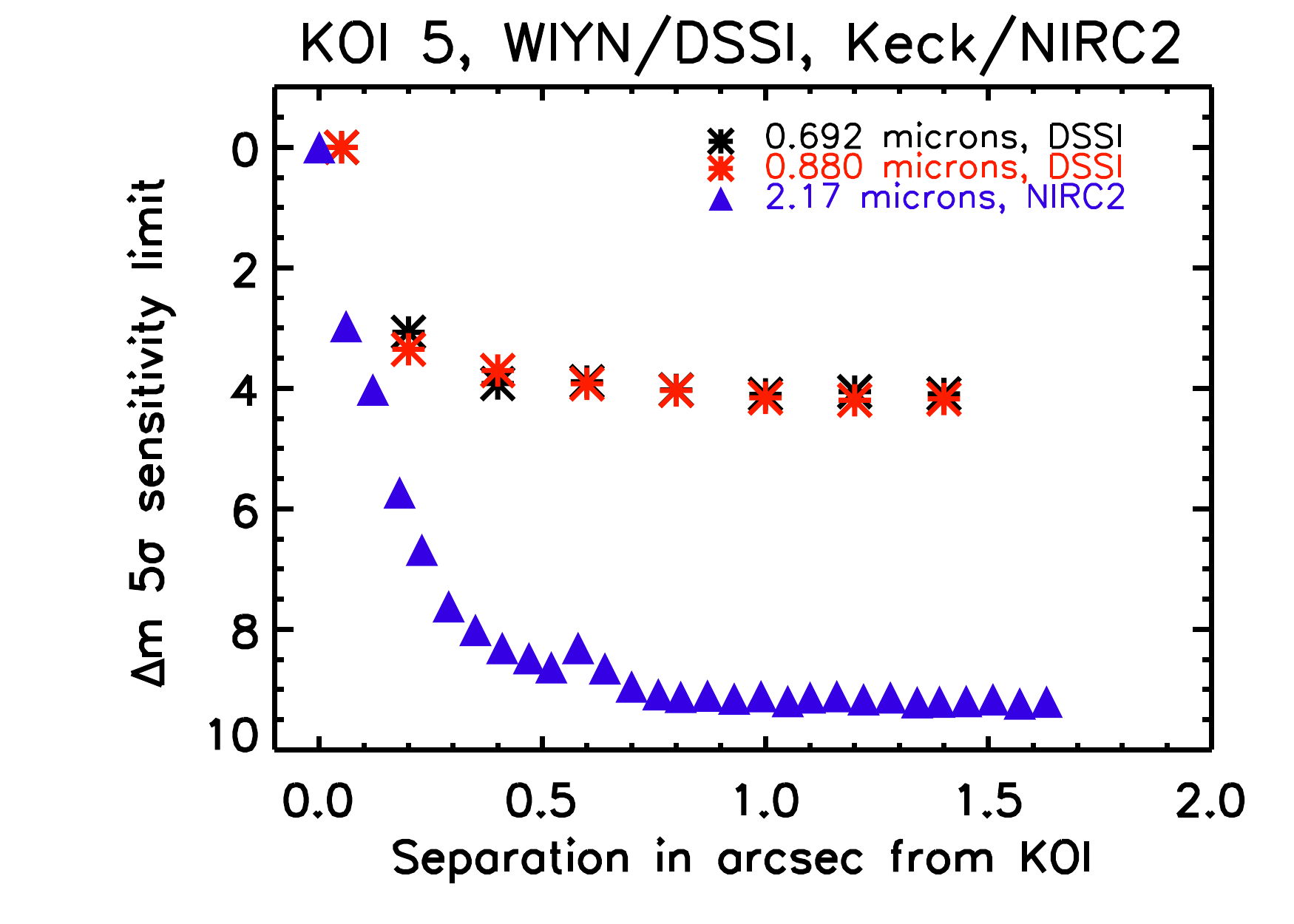}}
\hspace{-38pt}
\subfigure{\includegraphics[width=0.53\textwidth]{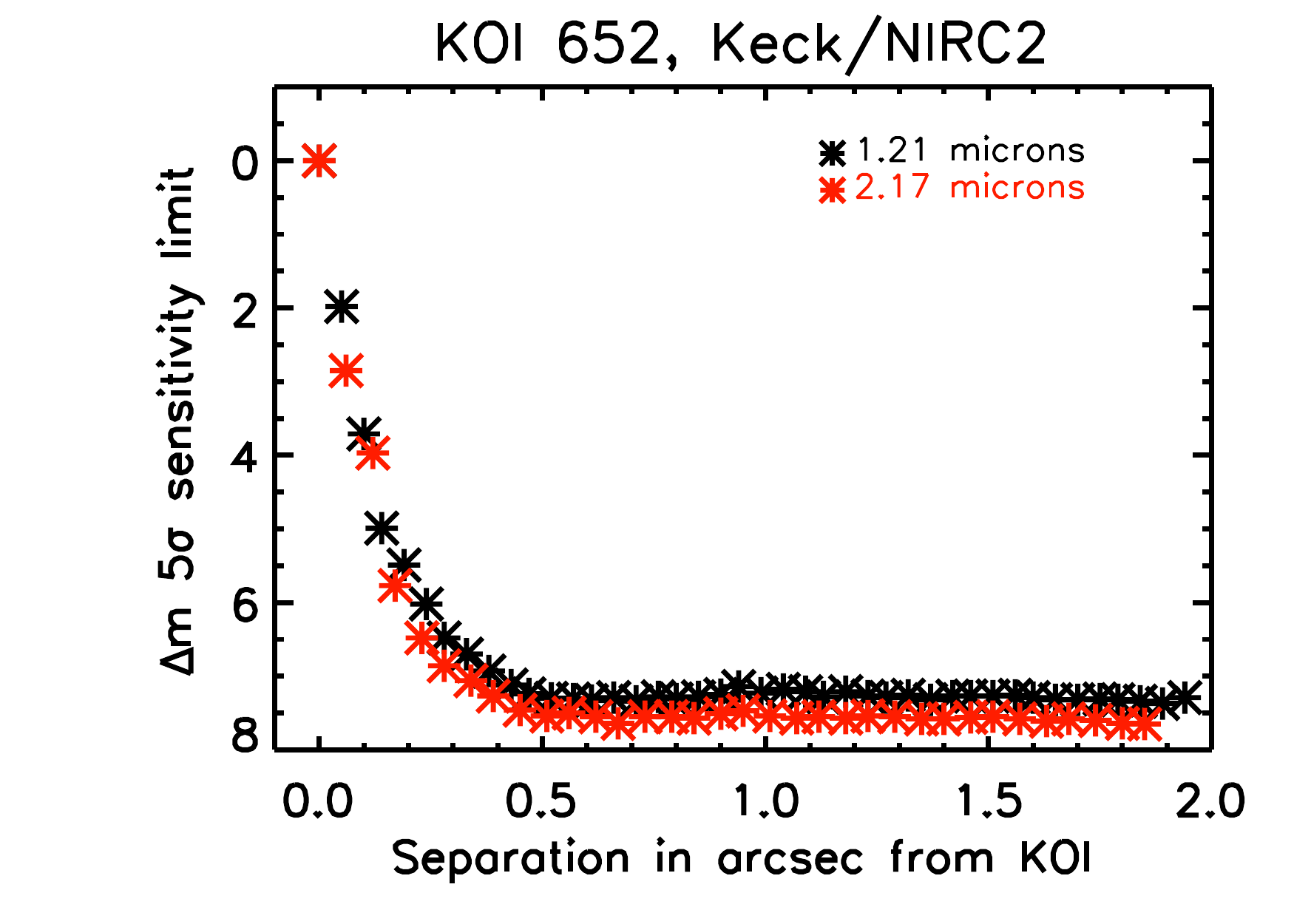}}
\hspace{-38pt}
\subfigure{\includegraphics[width=0.53\textwidth]{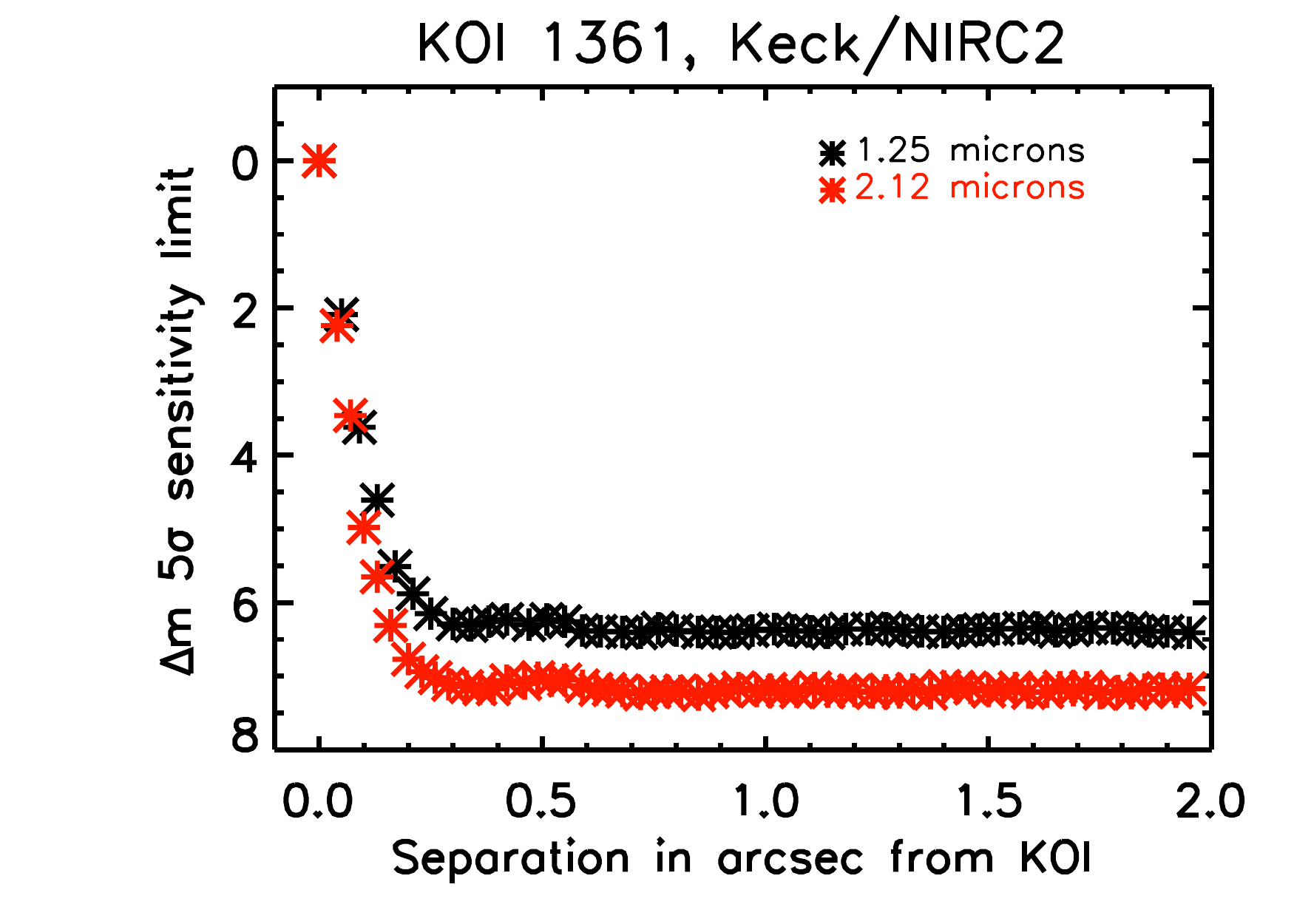}}
\hspace{-38pt}
\subfigure{\includegraphics[width=0.53\textwidth]{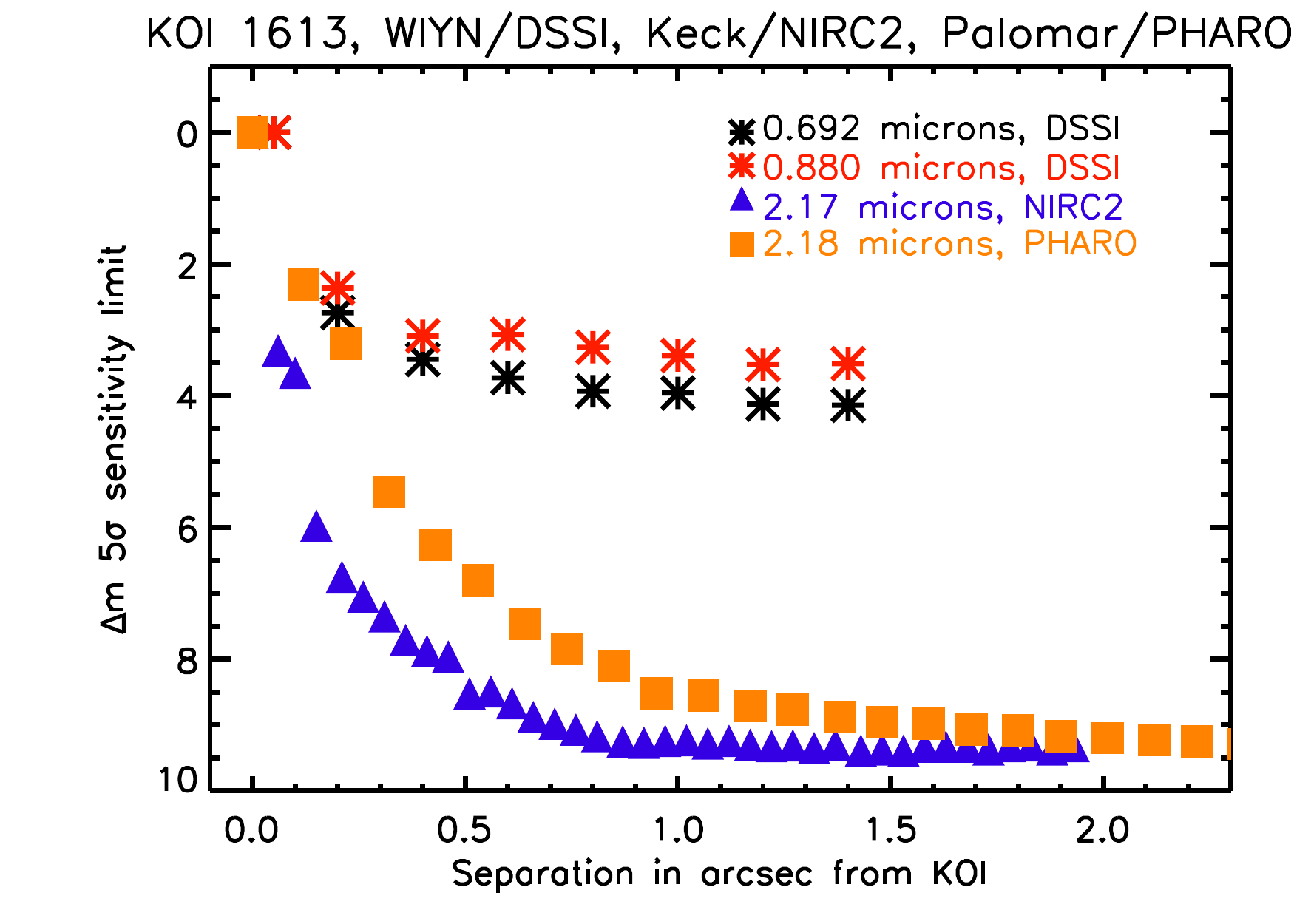}}
\hspace{-38pt}
\subfigure{\includegraphics[width=0.53\textwidth]{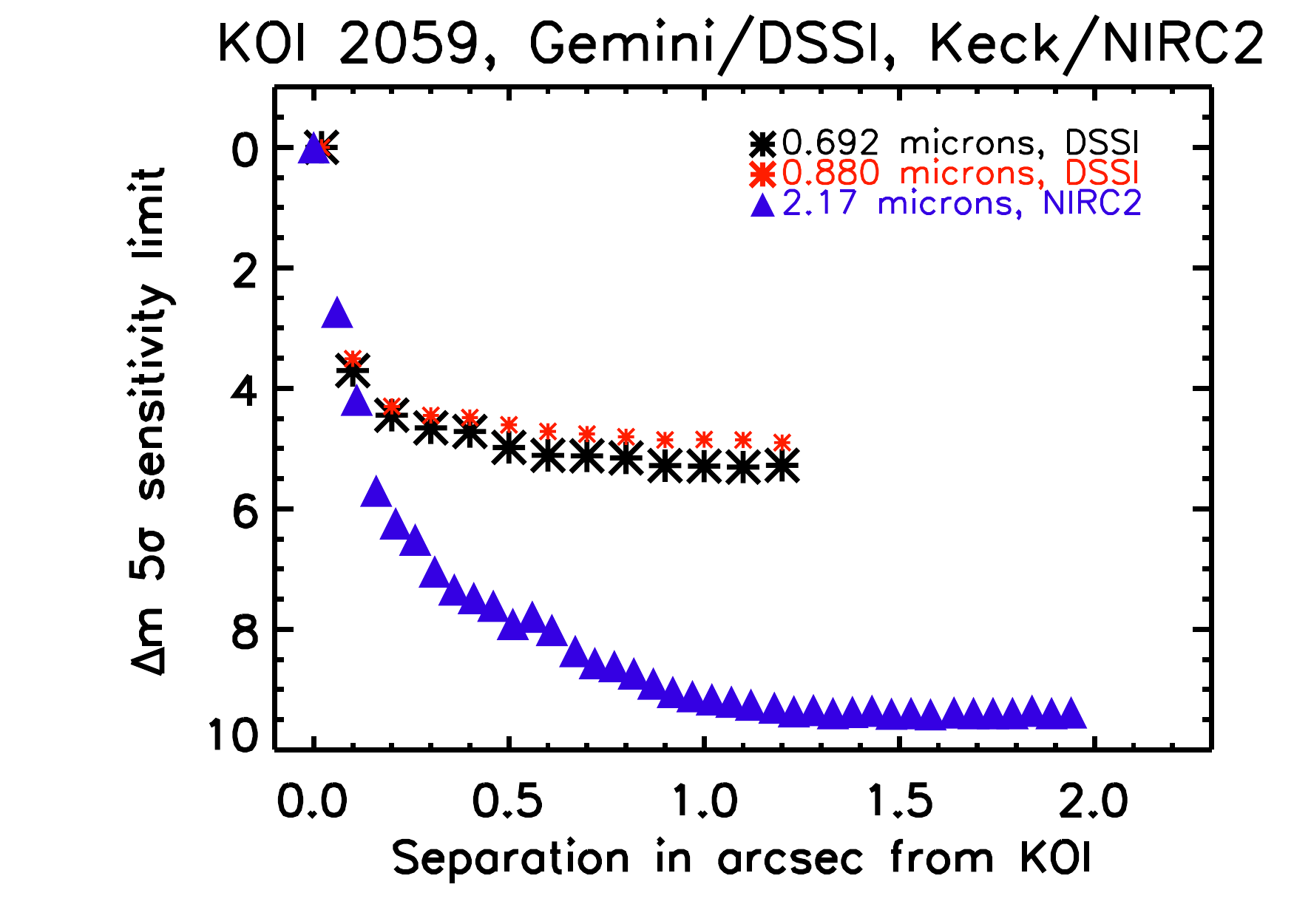}}
\hspace{-38pt}
\subfigure{\includegraphics[width=0.53\textwidth]{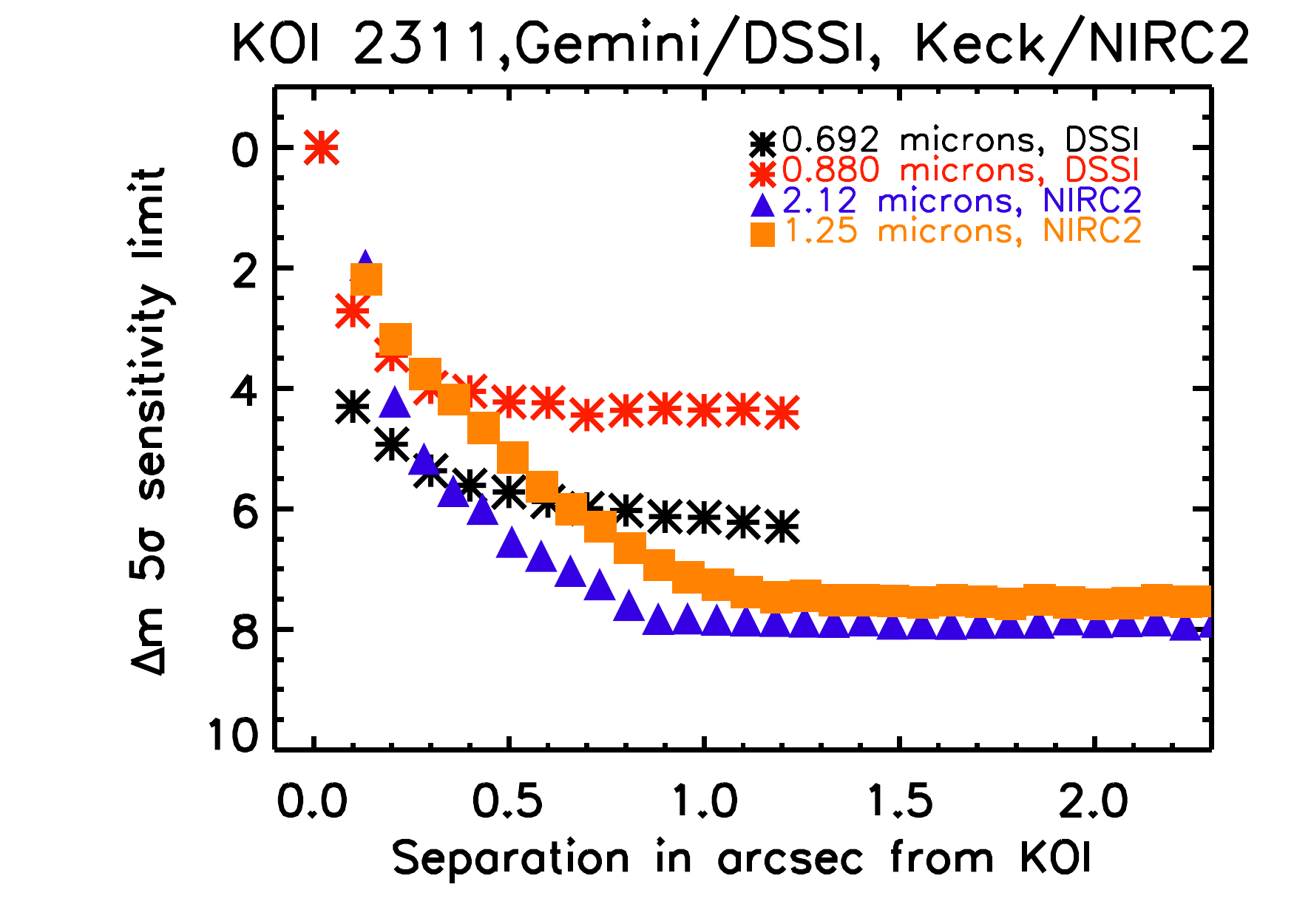}}
\caption{Curves depicting the 5$\sigma$ sensitivity limits of the
  imaging observations for each KOI in the K15 sample with a detection
in imaging data. Different colors and points correspond to different
wavelengths and instruments of the observations. }
\label{contrast_curves1}
\end{figure}

\begin{figure}[h]
\centering
\subfigure{\includegraphics[width=0.53\textwidth]{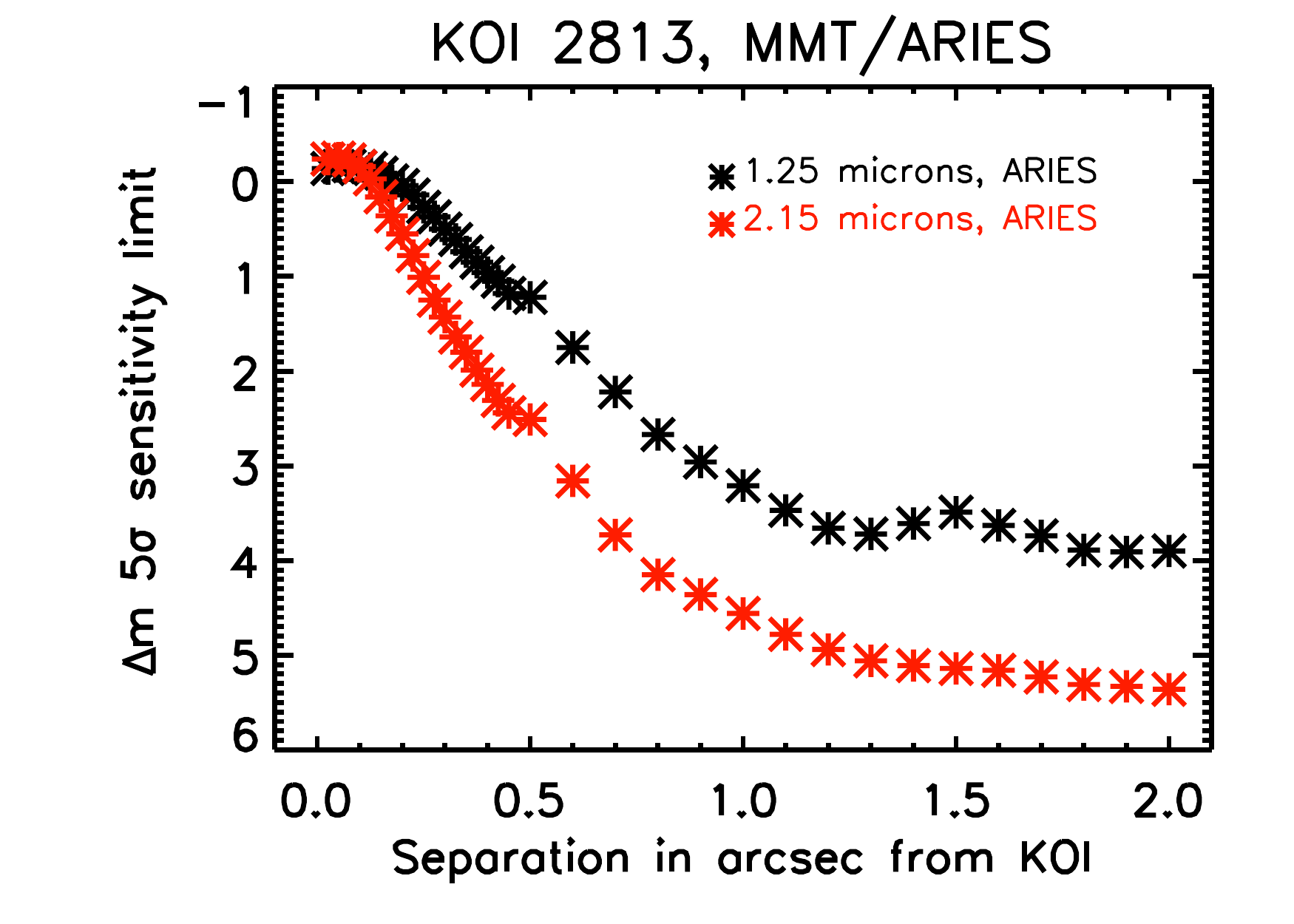}}
\hspace{-38pt}
\subfigure{\includegraphics[width=0.53\textwidth]{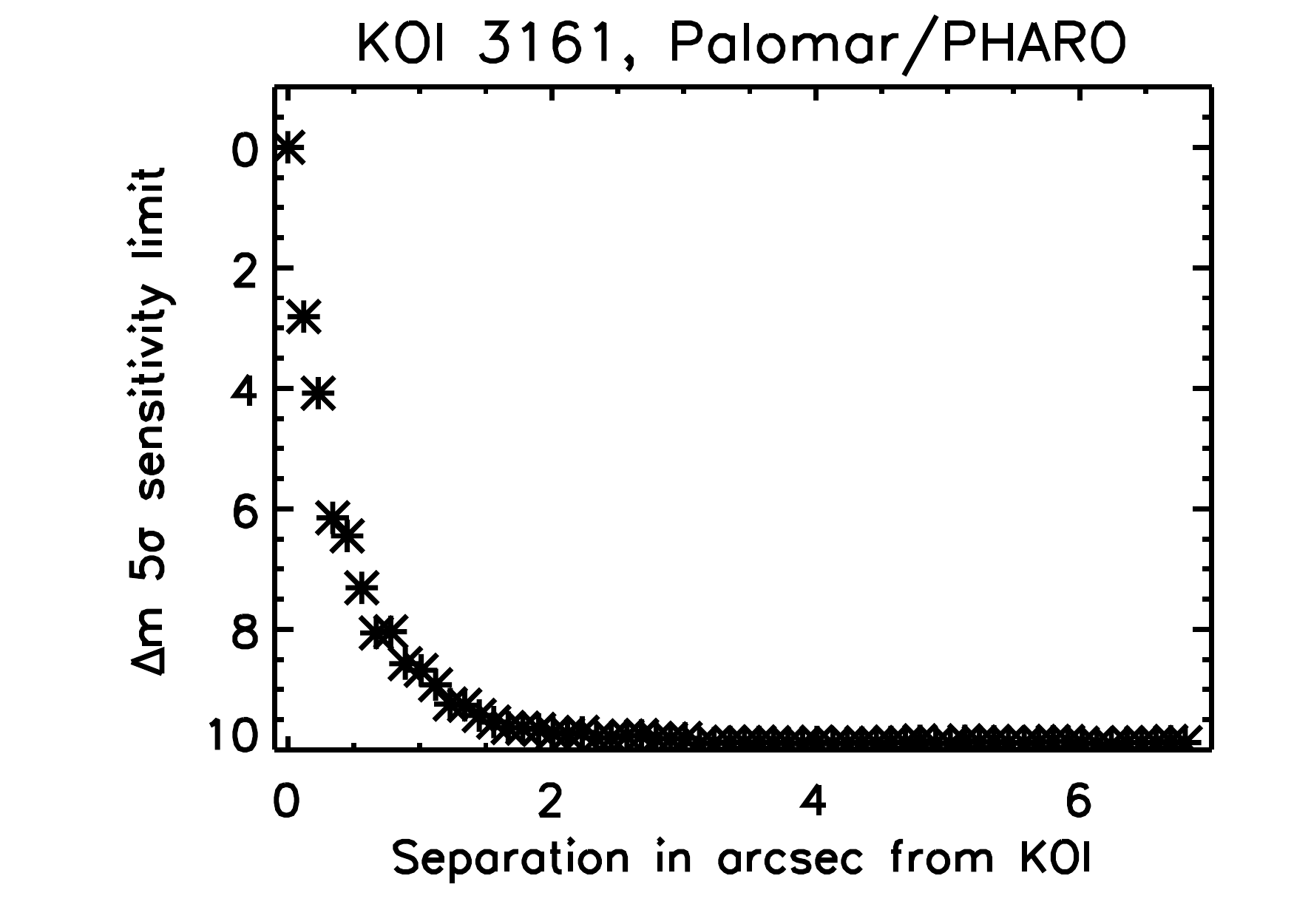}}
\hspace{-38pt}
\subfigure{\includegraphics[width=0.53\textwidth]{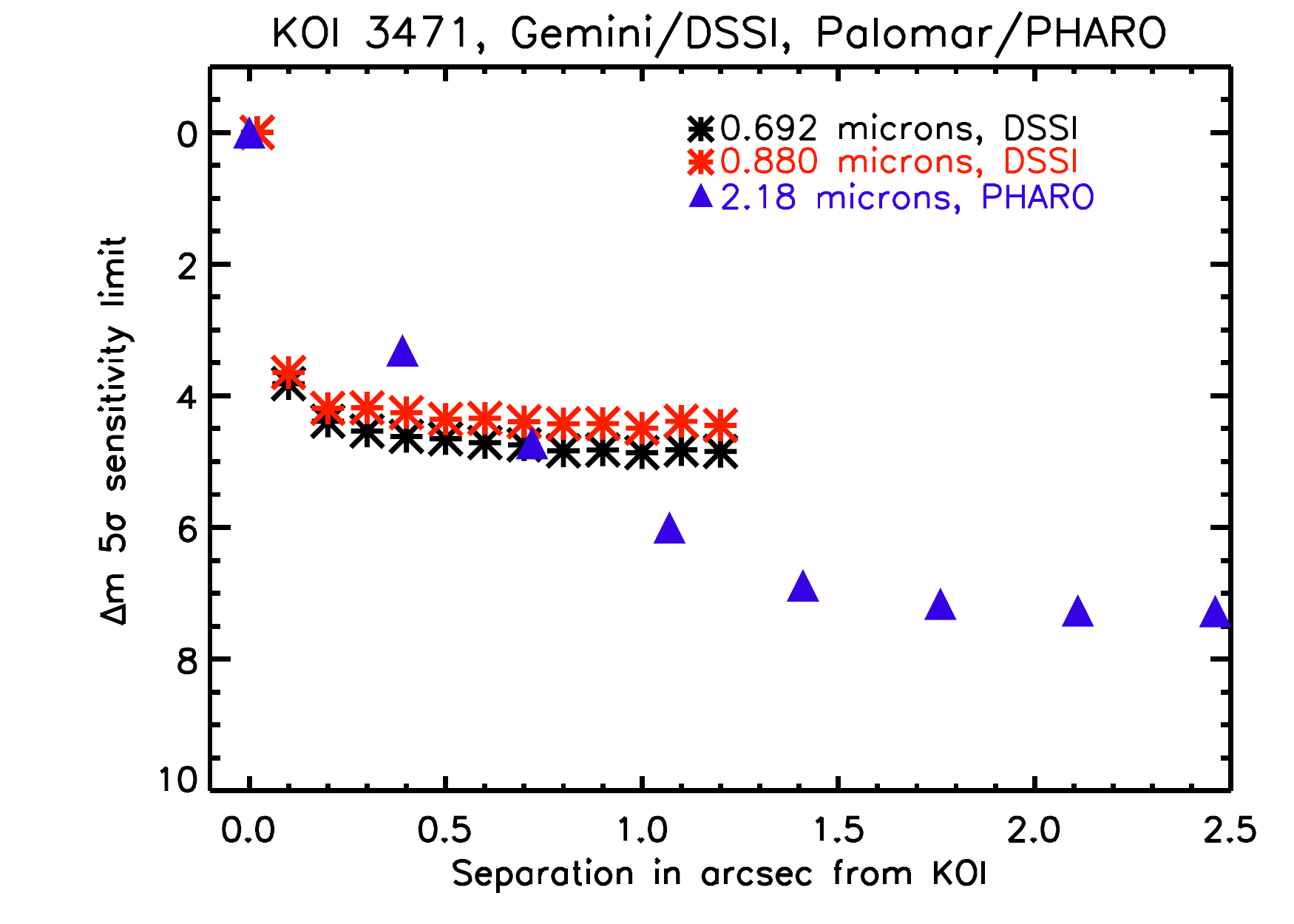}}
\hspace{-38pt}
\caption{Same as Figure \ref{contrast_curves1}.}
\label{contrast_curves2}
\end{figure}

Instead of examining the overlapping sample as a whole, we can
consider each KOI companion individually to try to pinpoint the cause of
differences between companion T$_{eff}$ values and flux ratios derived from
spectroscopy versus imaging analyses.  In the
following comparisons, the temperatures and flux ratios of the
companions detected and reported by K15 are compared to the companion
temperatures and flux ratios reported here, derived from imaging
observations, but we do not assume \textit{a priori} that the
companions are actually the same star. Included in the discussion
below are the limiting
magnitude contrast curves of each imaging observation, which are shown
for reference in Figures \ref{contrast_curves1} and
\ref{contrast_curves2}, the isochrone mapping for stars with more
than one color from \S3.1, and the separation of the
imaging-detected companions. We
assess whether, given all of the known information, the companions
detected via spectroscopy and imaging are likely to be the same or
different stars. 


\textbf{KOI 5 Companion} The flux ratio values overlap between methods
for this companion, and the T$_{eff}$ values are close 
when the errors are considered (Figure
\ref{secondary_comp}, top panel). The $\Delta$m and separation measured
from the \textit{WIYN/DSSI} imaging data put the companion near the limit of
detectability by imaging. The K15 $\Delta$m suggests their
companion would have been detected by the optical speckle imaging
outside of $\sim$0.1$\arcsec$, and detected by \textit{Keck II/NIRC2} if outside of
$\sim$0.07$\arcsec$ (Figure \ref{contrast_curves1}). The good agreement ($OVL$=0.677) in Figure \ref{iso_koi5} between the companion photometry contours, derived
assuming a bound companion, and the red point, derived from relative color
information and representative of the ``true'' color and relative
magnitude of the companion, points towards the imaging detection being
a bound companion. Furthermore, given the measured separation ($\sim$0.14$\arcsec$) of the
companion detected in \textit{WIYN/DSSI} images, the simulations of H14 predict a 94.2$^{+4.6}_{-10.2}$\% probability that the
companion is bound. In sum, the evidence slightly favors the spectroscopic and
imaging detections being the same star.   

\textbf{KOI 652 Companions} As mentioned above, K15 confidently reports three
companions around KOI 652 (and suggests a fourth), while imaging detects only two. Thus there
is some ambiguity as to exactly which parameters from the two methods to
compare. Here we assume the two hottest and highest-flux-ratio
detections from K15 are those most likely to correspond with the imaging
detections. Under this assumption, there is some agreement between the
T$_{eff}$ values and flux ratios derived by both methods -- one companion is
found here, based on imaging data, to be hotter than in K15
(4173$\pm$57~K from imaging vs. 3700$\pm$150~K from K15), and both are found in this work
to have higher flux ratios (0.35$\pm$0.10 and 0.17$\pm$0.05
 from imaging vs. 0.09$\pm$0.03 and 0.02$\pm$0.01 from K15). 
Interestingly, K15 note that
their method produces large systematic errors for companion stars with
T$_{eff}=$4000 K, mostly due to the sparsity of stellar models in their grid in
that temperature range. The separations measured from \textit{Keck II/NIRC2}
data for the companions to KOI 652 ($\sim$1.22$\arcsec$, 1.28$\arcsec$) are also some of the
largest in the sample, which may mean that if the imaging-detected companions are the same
as those detected spetroscopically by K15, their full flux did not fall
into the \textit{Keck I/HIRES} slit. Considering the flux ratios of
the K15-detected companions, in order of smallest to largest flux
ratio, they would be detectable by \textit{Keck II/NIRC2} if farther
than $\sim$0.2$\arcsec$, 0.17$\arcsec$, and 0.07$\arcsec$ in separation from KOI 652
(Figure \ref{contrast_curves1}). 
Both companions have photometry (red
points) that do \textit{not} overlap with the contours ``mapped'' down the
isochrone of the primary KOI assuming the stars are bound -- in all
cases the companions are too red (Figure \ref{iso_koi652}). This could be an artifact of the
Dartmouth isochrone extrapolation to such a cool, metal-poor KOI (see
Table 1) and its even cooler companions. However, the unbound nature of the companions
to KOI 652 is also supported by the simulations of H14, which indicate
that 14$^{+6.7}_{-5.2}$\% of companions detected at 692 nm with
\textit{Gemini/DSSI} beyond $>$1$\arcsec$ are bound. In this case the isochrone fit
plots in Figure \ref{iso_koi652}, the low probability of
the imaging-detected companions being bound, and the relatively large imaging companion
separations together suggest that the spectroscopic and imaging detections are \textit{not} of the same objects. 

\textbf{KOI 1152 Companion} In this case, the companion detected by
K15 seems to have a temperature (4200$\pm$350 K) close to the companion detected from the
imaging data according to our analysis in \S3.1 (3736$\pm$73 K), but the flux ratios
of the two data sets and analyses do not agree -- K15 finds F$_B$/F$_A=$0.31$\pm$0.14,
whereas our analysis finds F$_B$/F$_A=$0.80$\pm$0.29. The
imaging-detected companion's 0.59$\arcsec$ separation is within
the ideal separation range of K15, so the possibility that the
imaging-detected companion was just not fully in the \textit{Keck
  I/HIRES} slit for the K15 detection is smaller in this case than the
case of KOI 652's companions. However, as noted for KOI 652, this companion detection by K15 may be subject to
large systematic errors as K15 derives a T$_{eff}$ of
4200$\pm$350, within the range of sparsely sampled temperatures in
their stellar model grid. Furthermore, the K15-derived flux ratio is
higher (0.31$\pm$0.14) than their ``most accurate'' case of the companion star
contributing $\leq 20$\% of the total flux for the system. With
only one filter of imaging observations, no bound versus not-bound
analysis as in \S3.1 is possible in this case because the
companion's absolute photometry and true colors cannot be
determined. Given the separation of the companion, H14's simulation gives 94.2$^{+4.6}_{-10.2}$\%
probability with \textit{WIYN/DSSI} that
it is bound.  A companion
with the contrast ratio measured by K15 would be easily detectable by
\textit{Palomar/RoboAO}, according to the low performance contrast curves in Law et al. (2014),
so the reason for discrepancy between data/analyses is still uncertain,
but could be due to the acknowledged limitations of K15's method. 

\textbf{KOI 1361 Companion} In both temperature and flux ratio space
(Figure \ref{secondary_comp}, top panel), the parameters derived in
K15 and in this work are relatively consistent -- 3600$\pm$200 K
vs. 3262$\pm$22 K, and 0.02$\pm$0.01 vs. 0.04$\pm$0.01, respectively. According to the
\textit{Keck II/NIRC2} contrast curves (Figure \ref{contrast_curves1}), a companion with the
$\Delta$m derived by K15 would have to be closer than
$\sim$0.1$\arcsec$ to go undetected by the NIR AO imaging. The multi-color
imaging data from \textit{Keck II/NIRC2} plotted in Figure
\ref{iso_koi1361} shows overlap ($OVL$=0.869) between the relative
photometry contours of the companion, and overlap between
the relative photometry (red points) and that derived assuming the KOI and
companion are bound, although the errors are large; a wide range of
main sequence luminosities have similar J-K colors so the
discrimination between bound and field stars is not as strong from
this comparison. However, the 0.47$\arcsec$ separation of the
imaging-detected companion to KOI 1361
gives it a good chance of being a bound companion, according to H14's
\textit{Gemini/DSSI} simulation of 692 nm, which predicts a 71.1$^{+7.9}_{-8.9}$\% 
probability that a companion at 0.47$\arcsec$ is bound. Thus the cumulative
evidence suggests that the companion detected by spectroscopy
and imaging may indeed be the same star.

\textbf{KOI 1452 Companions} This system is a demonstration of
the necessity of imaging data, in addition to the spectroscopy
analysis of K15. The primary star is $\sim$7100 K, outside the range
of K15's search algorithm for companions in the primary KOI's
spectrum. K15 reports that the binarity in the spectrum is clear, but
they are unable to determine accurate parameters of the companion
star. From imaging data, two companions are detected, both $<$4000 K,
at 2.3$\arcsec$ and 4.8$\arcsec$ separations that would almost certainly have been missed by the
K15 method even if the primary star were cooler. At such wide
separations these stars are very unlikely to be bound to the primary KOI.

\textbf{KOI 1613 Companion} The K15-reported temperature
($>$6000$\pm$850 K) and flux
ratio ($>$0.04$\pm$0.01) of their detected companion are lower limits, which when combined with
their reported errors, could overlap with the
imaging-detected companion parameters in this work. K15 reports an RV separation of 10 km~s$^{-1}$ for
their detected companion, which is at the detectability limit of their
method. The multi-color imaging data in Figure \ref{iso_koi1613} indicate
marginal overlap between the relative photometry and the contours
derived assuming the companion is bound; in particular the 692-K
color is on the edge of agreement in 692 vs. 692-K space (third row of
Figure \ref{iso_koi1613}), and does not agree within 1$\sigma$ in K vs. 692-K space (fourth row). The error
on the 692-880 color is also large, such that the distinction between a bound
and an unbound companion is more ambigious than other cases. The
overlap coefficient ($OVL$) between the relative photometry contours (right-most
panels in Figure \ref{iso_koi1613})  = 0.845 for \textit{F692-F880} and 0.378
for \textit{F692-K}. The
simulations of H14 of \textit{WIYN/DSSI} data at 692 nm suggest a
$\sim$100\% probability that this companion, at $\sim$0.2$\arcsec$ separation,
is bound. Based on the contrast curves from imaging (Figure
\ref{contrast_curves1}), a
companion around KOI 1613 with the $\Delta$m found by K15 would have to be
closer than $\sim$0.4$\arcsec$ if measured by \textit{WIYN/DSSI}, $\sim$0.15$\arcsec$ if
measured by \textit{Palomar/PHARO}, and $\sim$0.05$\arcsec$ if measured by \textit{Keck
  II/NIRC2} to remain undetected. It is plausible that the K15
detection and the imaging detection are of the same companion, but as
the K15-detected companion parameters are only lower limits, and the
imaging-detected companions have large errors on their colors, 
this remains an ambiguous case. 


\textbf{KOI 2059 Companion} K15's reported RV separation between KOI
2059 and their detected companion is 5 km~s$^{-1}$, below their quoted detectability
limit for configurations other than a few specific cases (G-dwarf
primary+M-dwarf companion where the companion contributes $>$3\% of
the total flux). Thus the disagreement (see Figure \ref{secondary_comp}) between their companion T$_{eff}$ and flux
ratio and the same values derived in this work for the imaging-detected companions is perhaps
not surprising. According to the contrast curves in Fig \ref{contrast_curves1}, a companion with the flux ratio derived by K15 would need to be
within $\sim$0.4$\arcsec$ (visible)/0.14$\arcsec$ ($K$) to go undetected by imaging,
which is plausible. The measured separation of the imaging-detected
companion is 0.38$\arcsec$--0.39$\arcsec$ in $K$ band and the visible, so there
could yet be an inner companion detected spectroscopically by K15 and not by imaging. For $\Delta$RVs of $\leq 20$km~s$^{-1}$, and stars of
similar spectral type, K15 note that some of the light from their detected companion
may be subtracted along with that of the primary star,
thus causing the companion's flux to be underestimated. This could, alternatively,
be the case for the companion to KOI 2059; the imaging data analysis
in \S3.1 indicates a higher flux ratio (0.52$\pm$0.12) and higher companion T$_{eff}$
(4536$\pm$60 K) than in K15 (0.02$\pm$0.01, 3600$\pm$250 K). The lack of overlap between the contours and
red point in the panels of Figure \ref{iso_koi2059}, as well as the
minimal overlap ($OVL$=0.006) between the $\Delta$K and $\Delta$692 contours
(right-most panels in Figure \ref{iso_koi2059}), indicate that the
imaging-detected companion may be be unbound from the
KOI. 
However, based on the H14 simulations, the 0.39$\arcsec$ separation of
the imaging-detected companion to KOI 2059 suggests instead that it is
(94.2$^{+4.6}_{-10.2}$\%) likely to be bound. 

Interestingly, the $\sim$600 nm
\textit{Palomar/RoboAO} and $\sim$700 nm \textit{Gemini/DSSI} 
imaging data both suggest larger $\Delta$m values ($\sim$1) than the
$K$-band \textit{Keck II/NIRC2} imaging data ($\sim$0.12). Both the
visible and NIR imaging data find a separation of $\sim$0.38 and a
position angle of $\sim$290 deg, indicating the data are likely
targeting the same companion. This wavelength-dependent
$\Delta$m suggests a companion that is redder than the primary of the KOI
2059 system. K15's injection simulations of a 5500 K primary+3500 K companion
system with a $\Delta$RV of 5 km~s$^{-1}$ predict a recovery rate of
90\%, 90\%, and 40\% for companion star brightness fractions of 5\%,
3\%, and 1\%, respectively. The KOI 2059 primary is $\sim$5000$\pm$100 K, slightly
cooler than K15's simulation, but if the companion is an M dwarf,
unless it is $<$3\% the flux of the primary it has a good chance of
being detected by K15. Given the errors on the recovered parameter
uncertainties from K15's 5500 K primary+3500 K companion
system simulation (see their Table 6), the spectroscopic and imaging
companion detections may still be of the same object, but most
evidence indicates they are different stars. 

\textbf{KOI 2311 Companion} The parameters derived from the two
different methods significantly disagree for this companion, see
Figure \ref{secondary_comp}. The
imaging data indicate a separation of $\sim$1.03$\arcsec$, which is larger
than the likely upper separation limit noted by K15, but not outside
their hard separation limit of 1.5$\arcsec$. A companion with the flux ratio
measured by K15 ($>$0.28$\pm$0.07) should be detectable by imaging at almost any
separation (excluding $\lesssim$0.1$\arcsec$) as measured in the visible by
\textit{Gemini/DSSI} and NIR by \textit{Keck/NIRC2}. However, as noted by Everett et
al. (2015) and supported by the plots in Figure \ref{iso_koi2311}, the
imaged companion is very likely a faint background star. The absolute
photometry differs significantly from the ``assumed-bound'' case in
both 692-K and J-K color spaces, and
in 692-K color space the $\Delta$692 and $\Delta$K contours
(right-most panels) do not
overlap ($OVL$=0.0). Furthermore, H14's simulation of
\textit{Gemini/DSSI} observations at 692 nm suggest a 14.2$^{+6.7}_{-5.2}$\%
probability that a companion at this separation is bound. These lines
of evidence point towards the K15-detected companion being a different
star than the imaging-detected companion. 

\textbf{KOI 2813 Companion} The flux ratio measured for the companion
to KOI 2813 detected by K15 (0.20$\pm$0.06) versus the flux ratio of the companion
detected by Dressing et al. (2014) from \textit{MMT/ARIES} imaging
(0.11$\pm$0.04) are relatively consistent. However, the temperatures
derived by K15 and here in \S3.1 are quite discrepant, $>$6000$\pm$100 K
versus 3736$^{+59}_{-69}$ K. The K15 flux ratio
for this companion, F$_B$/F$_A=$0.195$\pm$0.06, puts its contribution at the limit of K15's ``most accurate'' case of the companion star
contributing $\leq 20$\%. A companion star with the flux ratio measured by
K15 would be undetected by MMT/ARIES if it were within $\sim$0.3$\arcsec$ as
measured in $K_s$ band or 0.6$\arcsec$ as measured in $J$ band, so perhaps this K15's detection is a very close
companion that is not detected in imaging data. The separation of the
\textit{MMT/ARIES}-detected companion, 1.04$\arcsec$, has a $\sim$20\% chance
of being bound, according to H14's \textit{Gemini/DSSI} simulations. 
In sum, it is difficult to say for certain whether the
companion detected by Dressing et al. (2014) is the same as detected
by K15. Follow-up with \textit{Keck II/NIRC2} and \textit{Gemini/DSSI} would provide a
smaller separation limit, and multiple colors that would allow a bound
analysis like that in \S3.1.

\textbf{KOI 3161 Companion} The separation of KOI 3161 from the
companion detected in \textit{Palomar/PHARO} observations is
$\sim$2.5$\arcsec$, beyond the hard detectability limit of K15. This suggests
that the two data sets and analysis methods detect different companions to KOI
3161, which is consistent with the large discrepancies between the two
methods in companion T$_{eff}$ and flux ratio. While K15 detects a hot
(lower limit T$_{eff} =$6000$\pm$100 K), relatively bright (F$_B$/F$_A =$0.31$\pm$0.09)
companion, our analysis of \textit{Palomar/PHARO} imaging observations
indicate a much cooler (3502$\pm$385 K), fainter (F$_B$/F$_A
=$0.002$\pm$0.003) companion. As expected, a companion with the flux
ratio derived by K15 should be detectable by NIR imaging unless it is
closer than $\sim$0.1$\arcsec$, which could be the case for the K15-detected
KOI 3161 companion. With
only one filter of imaging observations, no bound versus not-bound
analysis as in \S3.1 is possible in this case, but the large
separation (2.5$\arcsec$) of the imaging-detected companion makes it likely
to be unbound from KOI 3161 (H14). 

\textbf{KOI 3471 Companion} 
K15 reports three companions to KOI 3471, making it difficult to
compare with the single companion detected through imaging.  A further
complication is that the stellar properties of KOI 3471 listed by
Huber et al. (2014), which are based on broadband photometry, classify
KOI 3471 as a subgiant, but with a large uncertainty in log($g$) and
therefore ambiguity to the true luminosity class.  Some of the
difficulty classifying the star might be attributable to the blended
nature of its spectrum.

To understand the imaging results, we considered both a subgiant and a
dwarf scenario for this KOI by restricting the log($g$) values input to
the isochrone fit to 3.787$\pm$0.40 to represent a subgiant star and
4.6$\pm$0.20 to represent a dwarf (i.e., we adopt the Huber et al.
values for T$_{eff}$ and [Fe/H] and a wide range of log($g$) within each
luminosity class to accomodate the large uncertainty). Assuming a
subgiant primary, the isochrone fit results in $F_B/F_A =
0.058\pm0.068$ and T$_{eff}=4472\pm289$~K for the imaging-detected companion. With a
dwarf primary, we find $F_B/F_A = 0.035\pm0.012$ and
T$_{eff}=3470\pm36$~K. It is notable that for a subgiant primary, the observed color and
magnitude (red points in Fig \ref{iso_koi3471}, top two rows) disagree with the color and magnitude
extrapolated from the primary KOI assuming a bound companion; the
color observations reveal the imaging-detected companion is too red for its relative
faintness or too bright for its relative redness. The companion would
presumably be unbound in this scenario. However, assuming a dwarf
primary, the companion's photometry is in good agreement (including
$OVL$=0.716 for the \textit{F692-F880} relative photometry contours of the companion) with
expectations for a bound secondary, lending some credence to this
scenario.

The $T_{eff}$ and flux ratios of the three different companions found by K15
can be compared to the derived parameters of the imaged companion. The
flux ratios found by K15 
are all higher than the flux ratio derived from imaging;
 the F$_B$/F$_A =$0.75$\pm$0.08 value
reported by K15 is also well outside their ``most accurate'' case of the companion star
contributing $\leq 20$\%. However, both the flux ratio and
T$_{eff}$ for the faintest companion detected by K15 agree within
uncertainties with the properties derived from imaging in the case of
a subgiant primary. The K15 companion with
$F_A/F_B=0.20$ is too hot ($>6000$~K) in comparison to
the T$_{eff}=4472\pm289$~K (subgiant primary) or T$_{eff}=3470\pm36$~K
(dwarf primary) imaging-detected companion. 
The F$_B$/F$_A=0.75\pm0.08$ companion reported by K15 would be detected by both
\textit{Gemini/DSSI} and \textit{Palomar/PHARO} (see Figure
\ref{contrast_curves2}) unless within the FWHM of the image,
$\sim$0.02$\arcsec$ with \textit{Gemini/DSSI} and $\sim$0.1$\arcsec$ for \textit{Palomar/PHARO}. The
fainter companions reported by K15 would be detected outside of
$\sim$0.02$\arcsec$ and $\sim$0.1$\arcsec$ with \textit{Gemini/DSSI} and
\textit{Palomar/PHARO}, respectively. 

Thus, one of the K15-detected companions (the
highest or lowest flux ratio targets) may be the
same star as the imaging-detected companion -- the $F_A/F_B =
0.75\pm0.08$ K15 companion has a temperature consistent with the
imaging-detected companion parameters,
while the $F_A/F_B = 0.12\pm0.04$ K15 companion has a temperature and
flux ratio consistent with the imaging companion parameters, both assuming a
subgiant primary. The companion parameters derived from the imaging
data assuming a subgiant primary are not consistent with it being a
bound system (Fig \ref{iso_koi3471}), although H14 predicts a 94.2$^{+4.6}_{-10.2}$\%
probability that a companion at the 0.53$\arcsec$ separation of the imaging
detection is bound, giving weight to the dwarf primary
scenario. And yet, none of the K15 companions' parameters match the
imaging companion when the primary is assumed to be a dwarf. 

Slawson et al. (2011) identified this KOI as a 1000-day period eclipsing binary
with a relatively high contamination factor of 24\% (corresponding to
a flux ratio of $\sim$0.32 in K$_p$), although they
did not report a temperature, radius, or mass ratio for the binary
components. This period is inconsistent with the RV signal detected
by K15 
and inconsistent with the imaging-derived
flux ratios ($0.058\pm0.068$ in the subgiant primary case or $0.035\pm0.012$ in
the dwarf primary case). Overall, the agreement between the K15 and
imaging-detected companions is ambiguous, but it is likely that this
system has multiple companions (bound or not) to the primary
KOI.

\subsection{What are the Planet Radius Implications?}

As outlined in the introduction, if a stellar companion is responsible
for some fraction of the total flux, then the transit depth of the
planet will be diluted and the assumed radius of the planet will be
incorrect. The dilution factor depends on the radius of the star that
the planet transits -- whether it is the primary or a companion star
-- and the flux ratio of the primary to the companion star(s). 
The analysis in this paper does not address the host nature of the
KOIs (whether they or their companions host the planets), but we do
report and calculate flux ratios from K15 and imaging data. If we
assume that the primary stars (KOIs) are indeed the planet hosts, we
can calculate the planet radius increase factor as simply
$\sqrt{(F_{total})/(F_t)}$, where $F_t$ is the primary star that is
transited and F$_{total}$ is the total system flux, including any
companions. (Note that if the planet orbits the companion star, the
actual radius could be larger by factor of a few, i.e. larger than the
radii increases reported here.) This definition of planetary radius correction was
used by Ciardi et al. (2015) to estimate the average change in
\textit{Kepler}-detected planetary radii due to an undetected close
companion. Ciardi et al. (2015) found that if there are no follow-up
spectroscopic or imaging data, and KOIs are assumed to be single, that
on average the planetary radii may be underestimated by a factor of
1.5. This factor decreases to $\sim$1.2 if typical radial velocity and
high resolution imaging observations are available for the KOI, and is
also dependent upon the primary KOI spectral type (higher for earlier
type and lower for late type stars). 

\begin{figure}[h]
\centering
\includegraphics[width=1.0\textwidth]{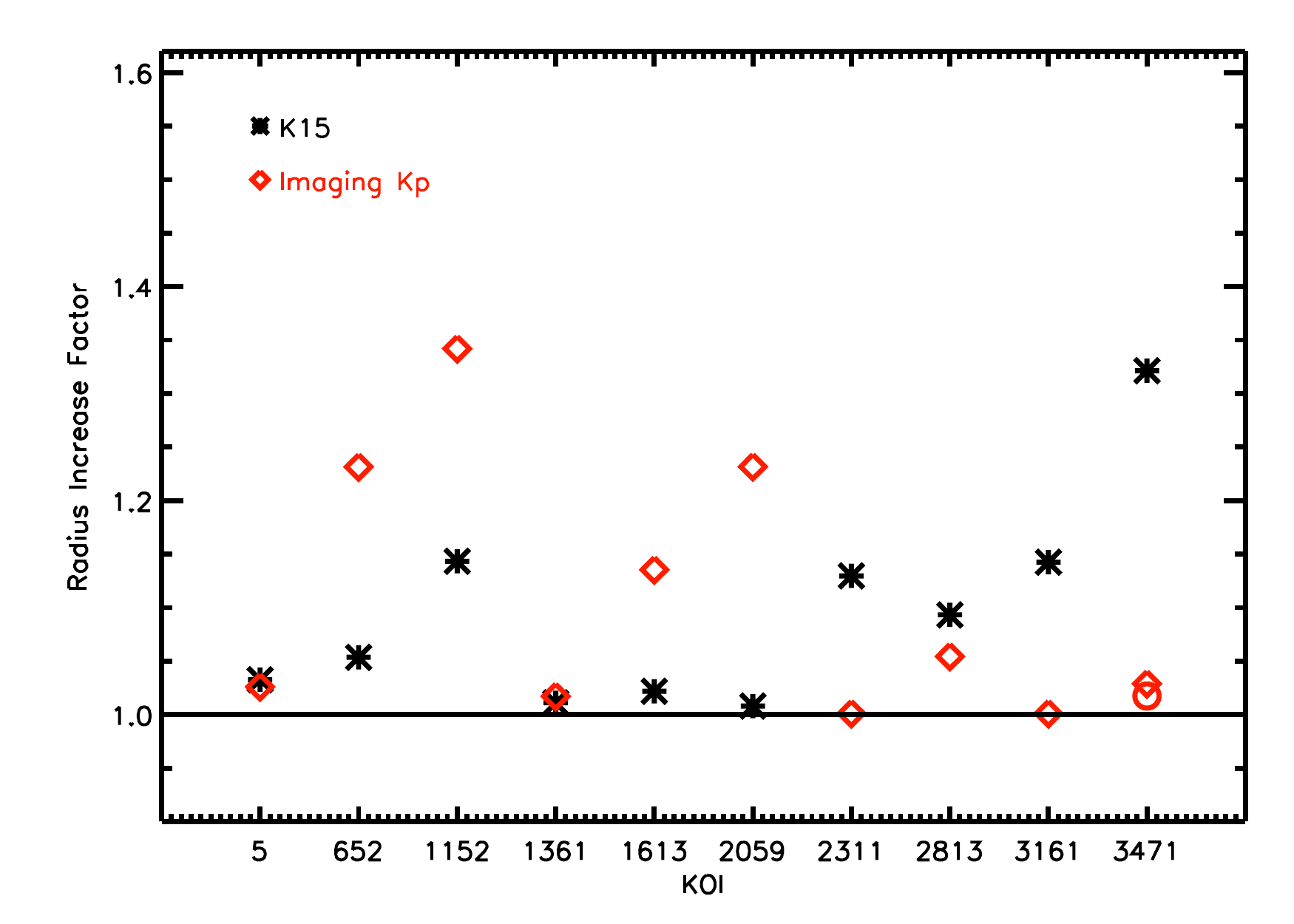}
\caption{Estimated increase in R$_p$/R$_{*}$ for each KOI when the
  detected companion is factored in, assuming the planet orbits the
  primary KOI. The radius increase factors derived from the K15
companion parameters are shown as black asterisks, and the radius
  increase factors from the imaging companion parameters derived in
 this work are shown as open red diamonds. The red circle above KOI
 3471 represents the radius increase from the detected companion,
 calculated under the assumption that the KOI is a dwarf (see \S4.2.2). A black horizontal line
  designates no radius increase.}
\label{pr}
\end{figure}

The resulting radius increase values for each of the
eleven systems considered in this work are plotted in Figure
\ref{pr}. The radii increase values based on the flux ratios of K15's
detected companions are
shown with black asterisks, while the values based on the companions
detected in imaging data and the companion parameter analysis
presented here are shown as red open diamonds (or, in the case of KOI
3471 being a dwarf, a red circle). In some instances, the
radius increase factor derived from the results of the two observations/techniques is very similar
(KOI 5, 1361, 2813), and in others the radius increase factors differ
substantially (KOI 652, 1152, 2059, 3471);
this is just another version of Figure \ref{secondary_comp}, right top
and bottom panels. The increase factors range from $\sim$1,
effectively no change in planet radius, to $\sim$1.3, which could
potentially change the status of a planet from ``rocky'' to
``non-rocky'', given the sharp transition radius of
$\sim$1.6 $\pm$0.02R$_{\oplus}$  (e.g., Marcy et al. 2014; Rogers
2015), e.g., KOI 2311.03 (1.44$\pm$0.16 R$_{\oplus}$). Furthermore, these predicted radius increase factors assume only one
companion around the KOI; if there are multiple companions (see \S5)
this would dilute the transit depth to a greater degree, making the
actual planet radius even larger. This
exercise illustrates how the discrepancies found in this paper can
manifest in broader exoplanet characterization and statistics, as also
shown by Ciardi et al. (2015).

\section{Summary and Conclusions}

The goal of this work was to investigate the overlap between companions to KOIs
found by two different techniques, deblending of high-resolution optical
spectroscopic observations versus high-contrast AO and/or speckle
imaging in the optical and NIR. 
Focusing on a sample of eleven stars that have companions detected
spectroscopically (K15) as well as companions detected through imaging
(this work, as well as other works listed in Table 1), we find few 
agreements -- 3/11 for companion T$_{eff}$ and 2/11 for F$_B$/F$_A$ --
but mostly disagreements between companion parameters derived from the
two sets of data and analysis methods. Examined as a whole, the
companion T$_{eff}$ values and F$_B$/F$_A$ ratios do not show an obvious pattern or dependence on
separation (as measured from the imaging data) from the primary KOI. Examined individually, and utilizing
contrast curves, isochrone ``mapping'' of the imaging observations,
and the measured separations of imaging-detected companions, the
differences between parameters from the two techniques can be explained
by: 

\begin{enumerate}

\item Limitations in both techniques to specific $\theta$ (separation) ranges
   -- too far away for spectroscopy or too close for imaging -- often
  meaning that the techniques actually detect different stars around
  each KOI

\item Limitations in the K15 parameter derivation method, such as the sparsity of
  cool stars in their stellar model grid, large uncertainty in derived
  parameters when the flux or spectral type of the
  primary KOI and the companion(s) are very similar, and $\Delta$RV
  constraints, as described in \S2.1.

\item The assumption that the companion is bound to the primary KOI 
  in the derivation of companion parameters from the imaging data -- in some cases this
    assumption is likely incorrect, resulting in spurious
    imaging-detected companion T$_{eff}$ values and F$_B$/F$_A$ ratios. This limitation is
    ameliorated by multi-color imaging of the companion, as described
    in \S3.1. 

\end{enumerate}

\noindent We summarize our findings and conclusions regarding
agreement between the two techniques for each KOI companion considered here in
Table \ref{tab3}. Based on our analysis, we can now help answer the following questions:

\noindent \textbf{Can spectroscopy find stars that imaging does not find?} Yes --
as expected, very close-in companions (e.g., $\theta \lesssim$0.02$\arcsec$-0.05$\arcsec$) are not easily
detectable with imaging. The different parameters of the companions
detected by K15 versus those detected in imaging data around KOIs
652, 2311, and 3161 are most likely explained by a small
separation for the K15-detected companions -- the imaging observations
are not detecting a very close-in companion that K15 detect. 

\noindent \textbf{Can imaging find stars that spectroscopy does not find?} Yes --
bound companions to KOIs at close separations ($\theta \ge $0.02-0.05$\arcsec$), companions that are
likely unbound at $>$1$\arcsec$ separation (e.g., companions to KOIs
652 and 2311), and/or
companions with small $\Delta$RV signals (e.g., KOIs 1613 and 2059), are difficult
to detect, and derive precise parameters for, using spectroscopic
deblending. Note that imaging data measures $\theta \sim$0.2$\arcsec$ for KOI
1613's companion, and $\theta \sim$0.4$\arcsec$ for KOI 2059's
companion; these would not be detected without high-contrast imaging
observations (they would not be detected in 2MASS data, for
example). Imaging observations are able to detect companions at
larger $\Delta$m and $\Delta$SpT than spectroscopy, and provide real measured
fluxes, position angles, and angular separations of companions. Due
to intrinsic and acknowledged limitations of the spectroscopic
deblending technique, it also does not perform well when the primary
star is off the Main Sequence (potentially KOI 3471), or has very cool
or very hot T$_{eff}$ (KOI 1452), no matter what the temperature or flux contribution of the
companion.  

\noindent \textbf{When are they likely to agree?} The techniques are most likely to
agree when the separation of the companion(s) is
$\theta \sim$0.1-0.8$\arcsec$, the secondary has F$_B$/F$_A \sim$1-20\%, both the primary and companion(s) are not too hot
(T$_{eff}>$6000 K) or too cool (T$_{eff}>$4000 K), and the $\Delta$RV between the
primary and companion is $>$10 km~s$^{-1}$.
Our work indicates that the
K15 spectroscopic deblending technique and imaging observations may be
detecting the same companion around
KOIs 5 and 1361. 

\noindent  There are four KOIs in our sample for which the spectroscopic
and imaging parameters do not agree, but the reason is not clear (KOIs
1152, 1613, and 2813, 3471). In the cases of KOIs 1152 and 2813, more colors of imaging data, and the
smaller separations probed by \textit{Keck II/NIRC2} or
\textit{Gemini/DSSI} data, would help
assess whether the imaging-detected companion is bound or not, and
thus help determine the likelihood of it being the same companion as detected
by K15. In the case of KOI 1613, more wavelength coverage of imaging
data may help better constrain the imaging-derived companion
parameters, but the K15 technique is limited by the system's
$\Delta$RV and, at present, can only provide upper limits on T$_{eff}$
and F$_B$/F$_A$. In the case of KOI 3471, the ambiguity in the primary
star parameters, the blended nature of the primary, and the multiple
K15-detected companions make a meaningful comparison with the imaging results challenging.


The spectroscopic deblending technique for detecting close-in
companions to KOIs, described in detail in K15, may find companions at
smaller separations than high-contrast imaging. However, the derived
properties from this method are often uncertain, and the method is
limited in the types of stars and companion configurations it can
detect, as well as the information it can provide. Thus, high-contrast
AO and speckle imaging provide an important complement, detecting a
wider range of companion types, at a larger range of separations,
around fainter stars. Our study illustrates why both techniques are
needed to fully characterize KOI multiplicity and
contamination, and can be used to test models of binaries in the
\textit{Kepler} field that could help better predict the number of
undetected binary host stars. Our study also shows the utility in combining
techniques -- in several cases (KOIs 1152, 1613, 2059, 2813,  3161,
3471) the combination of both techniques may indicate possible triple
or higher-order multiple
systems, with one companion detected with imaging observations and
one/more companion(s) detected spectroscopically. More astrometry,
wheather form speckle (common proper motions) or \textit{Gaia} could
help distinguish these cases.

\acknowledgements
The authors acknowledge the support of many people and
programs that made this work possible. This paper includes data
collected by the \textit{Kepler} Mission. Funding for the mission is
provided by the NASA Science Mission directorate. Most of the data presented here is made available to the
community for download at the \textit{Kepler} CFOP\footnote{https://cfop.ipac.caltech.edu}, a service of the
NASA Exoplanet Archive. These data include imaging-based separations
and $\Delta$m values, tabulated
sensitivity curves for each of the speckle observations, and KOI
stellar parameters.

M. E. Everett received support through NASA Agreement NNX1-3AB60A.

The \textit{WIYN} speckle imaging data presented here were based on
observations at Kitt Peak National Observatory, National Optical
Astronomy Observatory (NOAO 2010B-0241, 2011A-0130, 2013B-0115; PI: Howell), which is operated by the Association of Universities for Research in Astronomy (AURA) under a cooperative agreement with the National Science Foundation.

The \textit{Gemini} speckle imaging observations were obtained as part of
the programs GN-2013B-Q-87 and GN-2014B-Q-21 (PI: Howell) at the Gemini Observatory,
which is operated by the 
Association of Universities for Research in Astronomy, Inc., under a cooperative agreement 
with the NSF on behalf of the Gemini partnership: the National Science Foundation 
(United States), the National Research Council (Canada), CONICYT (Chile), the Australian 
Research Council (Australia), Minist\'{e}rio da Ci\^{e}ncia, Tecnologia e Inova\c{c}\~{a}o 
(Brazil) and Ministerio de Ciencia, Tecnolog\'{i}a e Innovaci\'{o}n Productiva (Argentina). We are very grateful for
the excellent support of the Gemin  administration and support
staff who helped make the visiting instrument program possible
and the DSSI observing run a great success.

Some of the data presented herein were obtained at the W.M. Keck
Observatory, which is operated as a scientific partnership among the
California Institute of Technology, the University of California and
the National Aeronautics and Space Administration. The Observatory was
made possible by the generous financial support of the W.M. Keck
Foundation. 

The Robo-AO system is supported by collaborating
partner institutions, the California Institute of Tech-
nology and the Inter-University Centre for Astronomy
and Astrophysics, and by the National Science Founda-
tion under Grant Nos. AST-0906060, AST-0960343, and
AST-1207891, by the Mount Cuba Astronomical Foun-
dation, by a gift from Samuel Oschin.

We thank the referee for their thoughtful comments and edits that improved the paper. 

Finally, the authors wish to recognize and acknowledge the very significant
cultural role and reverence that the summit of Mauna Kea has always
had within the indigenous Hawaiian community. We are most fortunate to
have the opportunity to conduct observations from this mountain.

{\it Facilities:} \facility{Gemini:Gillett (DSSI)}
\facility{Keck:II (NIRC2)}
\facility{PO:1.5m (Robo-AO)} \facility{MMT (ARIES)} \facility{CAO:2.2m (Astralux)}
\facility{WIYN:0.9m (DSSI)}

\begin{landscape}
\begin{deluxetable}{lcccccccc}
\tabletypesize{\scriptsize}
\tablecolumns{9}
\tablewidth{0pc}
\tablecaption{KOIs with K15 Companion Detections and Imaging Observations\label{tab1}}
\tablehead{ 
\colhead{KOI} &\colhead{KIC ID} & \colhead{Component} & \colhead{ d
  ($\arcsec$) }  & \colhead{KOI Mag ($K_p$)}& \colhead{$\Delta$Mag in
Imaging Filter}& 
\colhead{Imaging Band/Filter}  & \colhead{Source} & \colhead{Instrument} }
\startdata
5 	&   8554498  	& 	B	& 	0.136	&  11.67&	2.305$\pm$0.021	&	K	&	this work	&	Keck/NIRC2\\
5 	&   8554498  	& 	       B          	& 	0.142	&  11.67	& 	2.88$\pm$0.15	&	562	&	this work	&	WIYN/DSSI\\
5 	&   8554498  	& 	        B         	&   0.142	    	&  11.67    & 3.04	$\pm$0.15 &692 &	this work	&	WIYN/DSSI\\
652 &   5796675	& 	B	& 	1.222	& 13.65	&0.993$\pm$0.033	&	J	&	this work	&	Keck/NIRC2\\
652	&   5796675& 		       C         & 	1.283 & 13.65	& 	1.662$\pm$	0.034	&	J	&	this work	&	Keck/NIRC2\\
652	&   5796675& 		      B         & 	1.221 & 13.65	& 	0.62	$\pm$	0.03	&	K	&	this work	&	Keck/NIRC2\\
652 &   5796675	& 	C	              & 	1.283 & 13.65	& 	1.28	$\pm$	0.03	&	K	&	this work	&	Keck/NIRC2\\
1152 & 10287248& 	B	& 	0.59	& 13.99 &	0.31	$\pm$	0.31&	LP600	&	Law et al. (2014)	&	Palomar/Robo-AO\\
1361 &6960913& 	B	& 	0.467	& 15.00	&2.872$\pm$0.034	&	J	&	this work	&	Keck/NIRC2\\
1361 &6960913& 	B	                & 	0.474 &15.00	& 	2.884	$\pm$0.028	&	K	&	this work	&	Keck/NIRC2\\
1452 &7449844& 	B	& 	2.371	&  13.63&	9.284$\pm$0.9354&	i	&	Lillo-Box et al. (2014)	&	CalarAlto/Astralux\\
1452 &7449844& 	C	                & 	4.763 &13.63	& 	5.953$\pm$0.361	&	i	&	Lillo-Box et al. (2014)	&	CalarAlto/Astralux\\
1613 &6268648& 	B	& 	0.209	&  11.05&	0.857	$\pm$	0.036	&	K	&	this work	&	Keck/NIRC2\\
1613 &6268648& 	B	       & 	0.22	& 11.05&	1.3	$\pm$	0.22	&	i	&Law et al. (2014)	&	Palomar/Robo-AO\\
1613 &6268648& 	 B             	& 	0.212 &11.05	& 	1.28	$\pm$	0.15	&	692	&	this work	&	WIYN/DSSI\\
1613 &6268648& 	  B            	& 	0.207 &11.05	& 	1.28	$\pm$	0.15	&	880	&	this work	&	WIYN/DSSI\\
1613 &6268648& 	   B           	& 	0.19	& 	11.05&0.726	$\pm$	0.01	&	K	&	this work	&	Palomar/PHARO\\
2059 &12301181& 	B	& 	0.3866	& 12.91	&1.05	$\pm$	0.15&	692	&this work	&	Gemini/DSSI\\
2059 &12301181& 	      B        	& 	0.383 &12.91	& 	0.116	$\pm$	0.03	&	K	&	this work	&	Keck/NIRC2\\
2059 &12301181& 	      B        	& 	0.38	&  12.91&	1.1	$\pm$	0.14	&	LP600	&	Law et al. (2014)	&	Palomar/Robo-AO\\
2311 &4247991& 	B	& 	1.03	&  12.57 &	5.47	$\pm$	0.15&	692	&	Everett et al. (2015)	&	Gemini/DSSI\\
2311 &4247991& 	 B             	& 1.0264 & 12.57	& 	5.38	$\pm$	0.13&	J	&	Everett et al. (2015)	&	Keck/NIRC2\\
2311 &4247991& 	  B            	& 1.0264&	12.57	& 	4.74	$\pm$0.06&	K	&	Everett et al. (2015)	&	Keck/NIRC2\\
2813 &11197853& 	B	& 	1.04	& 13.59&	1.82	$\pm$	0.02&	Ks	&	Dressing et al. (2014)	&	MMT/ARIES\\
3161 &2696703& 	B	& 	2.546 &9.58	& 	4.213	$\pm$	0.025	&	K	&	this work	&	Palomar/PHARO\\
3471 &11875511& 	B	& 	0.532 &13.34	& 	3.7$\pm.23$&	692 	&this work&	Gemini/DSSI\\
3471 &11875511& 	      B                  	& 	0.527 &13.34	& 	2.81	$\pm$.13&	880 	& this work	&	Gemini/DSSI\\
\enddata
\label{tab1}
\end{deluxetable}
\end{landscape}

\begin{landscape}
\begin{deluxetable}{lcc|ccc||cc}
\tabletypesize{\scriptsize}
\tablecolumns{8}
\tablewidth{0pc}
\tablecaption{Parameters Derived from K15 Spectroscopy and Imaging (This Work)}
\tablehead{ 
\multicolumn{3}{c}{   } & \multicolumn{3}{c}{Spectroscopy Derived Parameters (K15)} & \multicolumn{2}{c}{Imaging Derived Parameters} \\
\colhead{KOI} &\colhead{KIC ID} & \colhead{Component} & \colhead{K15 F$_B$/F$_A$} &
\colhead{K15 $\Delta$RV (kms$^{-1}$) }& \colhead{K15 Comp T$_{{eff}}$ (K)} &
\colhead{Imaging Comp F$_B$/F$_A$ in $K_p$ $^{a}$}  & \colhead{Imaging Comp T$_{{eff}}$ (K) $^{a}$}
}
\startdata
5 	&   8554498  & B   &	$>$ 0.066$\pm$0.02&	11	& 5900$\pm$850 &	0.053$\pm$0.008	&	4564$\pm$55	\\
652	 &   5796675  & B  &	0.092	$\pm$	0.028	&	22&	3700	$\pm$	150	 & 	0.351	$\pm$	0.101$^{+}$	&	4173	$\pm$	57$^{+}$	\\
652	 &   5796675  & C &	0.02	$\pm$	0.007	&	46&	4000	$\pm$	350	&		 0.166	$\pm$	0.047$^{+}$	&	3962	$\pm$	50$^{+}$	\\
652	 &   5796675  & D  &	0.006	$\pm$	0.002	&	-44	&	3500	$\pm$	150	& 		\nodata		&		\nodata	\\
1152 & 10287248 & B	&	0.307	$\pm$	0.138	&	27	&	4200	$\pm$	350	&	0.801	$\pm$	0.294	&	3736	$\pm$	73	\\
1361 &6960913 & B	&	0.022	$\pm$	0.007	&	40	&	3600	$\pm$	200	&	0.035	$\pm$	0.011	&	3262	$\pm$	22	\\
1452 &7449844 & B	&	outside temp range			&	81 &	\nodata	$\pm$		& 	 0.0001	$\pm$ 0.00019	&	3263	$^{+42}_{-82}$	\\
1452 &7449844 & C	& outside temp range		&	81	&		\nodata	$\pm$		&		0.002		$\pm$0.004	&	3625 $^{+402}_{-110}$		\\
1613 &6268648 & B	&	$>$ 0.044	$\pm$	0.013	&	10	&	$>$ 6000 	$\pm$	850	&		0.289	$\pm$	0.046	&	5553	$\pm$	57	\\
2059 &12301181 & B	&	0.016	$\pm$	0.008	&	5	&	3600	$\pm$	250	&		0.517	$\pm$	0.124$^{+}$	&	4536	$\pm$	60$^{+}$	\\
2311 &4247991	& B&	$>$ 0.276	$\pm$	0.069	&	11	&	5600	$\pm$	400	&		0.002	$\pm$	0.0004$^{+}$	&	3288	$\pm$	19$^{+}$	\\
2813 &11197853 & B	&	0.195	$\pm$	0.058	&	26		&	$>$ 6000 	$\pm$	100	&		0.112$\pm$ 0.037	&	3736$^{+59}_{-69}$		\\
3161 &2696703 & B	&	0.305	$\pm$	0.092	&	-167&	$>$ 6000 	$\pm$	100	&		0.002 $\pm$0.003	&	3502$^{+385}_{-109}$	\\
3471 &11875511 & B&	0.746	$\pm$	0.079	&	-33		&4600	$\pm$	200	&	 0.058$\pm$0.068$^{*}$/0.035$\pm$0.012  &4472$\pm$289$^{*}$/3470$\pm$36  \\
3471 &11875511 & C	&	0.202	$\pm$	0.061	&	26	&	$>$6000	$\pm$	100	&		\nodata		&	\nodata	\\
3471 &11875511 & D	&	0.117	$\pm$	0.041	&	-52	&	4100	$\pm$	300	&	\nodata		&	\nodata		\\
\enddata
\label{tab2}
\tablecomments{``Comp'' columns indicate companion or ``B'' componenet
  of system. $^{a}$These values are based on the ``isochrone-shifted'' analysis of the imaging data, which is
possible for all the KOIs, versus multi-color differential photometry
analysis, which is impossible for KOIs with only one color of imaging
data (see Table \ref{tab1}). The flux ratio is measured from
\textit{Kepler} bandpass ($K_p$) fluxes. These temperatures and their errors are derived from the
contours in Figures 3-9. A $^{+}$ symbol indicates the companion is
likely to be unbound based on our analysis in \S3.1. A $^{*}$ symbol indicates parameters based on a fit to
the imaging companion measurements assuming the primary is a subgiant,
versus a dwarf (no-$^{*}$ values).}
\end{deluxetable}
\end{landscape}

\begin{landscape}
\begin{deluxetable}{lccccc}
\tabletypesize{\scriptsize}
\tablecolumns{6}
\tablewidth{0pc}
\tablecaption{}
\tablehead{ 
\colhead{Companion to KOI} & \colhead{T$_{eff}$s}&\colhead{F$_B$/F$_A$s} & \colhead{Bound according to} & \colhead{Bound
  according to} & \colhead{Likely same or} \\
\colhead{} & \colhead{consistent?} & \colhead{consistent?} &
\colhead{photometric+isochrone analysis?} & \colhead{Horch et al. (2014) comparison?} & \colhead{different companion?}}
\startdata
5 & no & yes & yes & yes & same \\
652B & almost & no & no & no & different -- 1, 2, 3\\
652C & yes & almost & no & no & different -- 1, 2, 3 \\
1152 & almost & no & \nodata & yes & uncertain -- 2 \\
1361 & almost & yes & yes & yes & same \\
1452 & \nodata & \nodata & \nodata & no & uncertain -- 1, 2, 3\\
1613 & yes & no & maybe & yes & uncertain -- 2, 3\\
2059 & no & no & no & yes & different -- 1, 2, 3\\
2311 & no & no & no & no & different -- 1, 3\\ 
2813 & no & almost & no & no & uncertain -- 1, 3\\
3161 & no & no & \nodata & no & different -- 1, 3 \\
3471 (subgiant primary) & yes & no & no & yes & uncertain \\
3471 (dwarf primary) & no & no & yes & yes  & uncertain \\
\enddata
\label{tab3}
\tablecomments{In the last column, the likely reasons for the
  discrepancies between spectroscopic and imaging detected companions
  are listed as numbers corresponding to the three reasons listed in
  \S5. The companion to KOI 3471 is uncertain because the parameters
  of the primary are uncertain.}
\end{deluxetable}
\end{landscape}

\end{document}